\documentclass[useAMS,usenatbib,usegraphicsx,fleqn]{mn2e}
\usepackage{graphicx}
\usepackage{natbib}
\usepackage{url}
\usepackage{color}
\usepackage{times}
\usepackage{pdflscape}

\title[Near-infrared spectroscopy of Massive Young Stellar Objects]{Medium resolution near-infrared spectroscopy of Massive Young Stellar Objects}
\author[R. Pomohaci et al.]{R. Pomohaci\thanks{E-mail:
pyrp@leeds.ac.uk}, R. D. Oudmaijer, S. L. Lumsden, M. G. Hoare and I. Mendigut\'ia
\\
School of Physics and Astronomy, The University of Leeds, Woodhouse Lane, Leeds LS2 9JT, United Kingdom.\\
}
\begin{document}

\pagerange{\pageref{firstpage}--\pageref{lastpage}} \pubyear{2015}

\maketitle

\label{firstpage}

\begin{abstract}
  
We present medium-resolution (R$\sim$7000) near-infrared echelle
spectroscopic data for 36 MYSOs drawn from the Red MSX Source (RMS)
survey.  This is the largest sample observed at this resolution at
these wavelengths of MYSOs to date.  The spectra are characterized
mostly by emission from hydrogen recombination lines and accretion
diagnostic lines. One MYSO shows photospheric HI absorption, a
comparison with spectral standards indicates that the star is an A
type star with a low surface gravity, implying that the MYSOs are
probably swollen, as also suggested by evolutionary calculations. An
investigation of the Br$\gamma$ line profiles shows that most are in
pure emission, while 13$\pm$5\% display P Cygni profiles, indicative
of outflow, while less than 8$\pm 4$\% have inverse P Cygni profiles,
indicative of infall. These values are comparable with investigations
into the optically bright Herbig Be stars, but not with those of
Herbig Ae and T Tauri stars, consistent with the notion that the more
massive stars undergo accretion in a different fashion than lower
mass objects which are undergoing magnetospheric accretion. Accretion
luminosities and rates as derived from the Br$\gamma$ line
luminosities agree with results for lower mass sources, providing
tentative evidence for massive star formation theories based on
scaling of low-mass scenarios.  We present Br$\gamma$/Br12 line
profile ratios exploiting the fact that optical depth effects can be
traced as a function of Doppler shift across the lines. These show
that the winds of MYSOs in this sample are nearly equally split
between constant, accelerating, and decelerating velocity
structures. There are no trends between the types of features we see
and bolometric luminosities or near-infrared colours.

\end{abstract}


\section{Introduction}
\label{sec:intro}

Understanding the star formation process is of crucial importance for
a number of branches of astrophysics. However, there are still gaps in
our knowledge of it. Low-mass star formation is thought to result from
the monolithic collapse of a cloud followed by accretion through the
circumstellar disk, and is in general reasonably well understood as
per the description of \citet{1987ARA&A..25...23S}.  The mechanism
through which the low mass pre-Main Sequence stars (of order few solar
masses and less) further grow via this accretion disk is widely
accepted to be due to magnetically controlled accretion. The infalling
material flows along magnetic field lines onto the proto-star
\citep{bertout1989}. This paradigm has been tried and
observationally tested in the case of low mass T Tauri stars.  A review
describing much of the evidence is provided by \citet{bouvier2007}.

The theory needs to be adapted if it is to be applied to the formation
of objects that are more massive. Early calculations by
\citet{1974A&A....37..149K} showed that the radiation pressure can
halt accretion for objects 20 M$_{\odot}$ and larger under spherical
accretion, essentially putting an upper limit to the mass of a
star. However, this mass is well below that observed from direct
measurements, and theoretical efforts now converged to two main
scenarios. Studies by \citet{2009Sci...323..754K} have shown that a 40
M$_{\odot}$ star can be formed by the monolithic collapse of a dense
cloud followed by disk accretion, with radiation pressure escaping
along bipolar outflow cavities. A different approach has been taken by
\citet{1998MNRAS.298...93B} and \citet{2006MNRAS.370..488B}. They
propose competitive accretion as a solution to the high-mass star
formation problem, whereby the most massive stars are formed at the
center of dense stellar clusters. The gravitational potential of the
massive star causes the gas to be funneled toward it instead of the
other proto-stars, allowing it to grow quicker and to higher mass. In
this scenario, a circumstellar disk is also required.

Although disk accretion seems therefore to be a main contender for the
formation of massive stars, little is known about the precise
accretion mechanism.  Even the most recent, sophisticated star
formation models are not able to simulate the fine detail required to
probe the accretion process from parsec scales via the disk to the
stellar surface. Indeed, models use a (large) volume as a sink
particle in which matter falls into and assume this material will
continue to form the star. For example, \citet{Rosen2016} explicitly
mention that the material is not followed in the inner 80 au.
Furthermore, at present, it is not at all known how the disk accretion
would operate in massive stars. Once in their Main Sequence
configuration, these stars are generally non-magnetic due to their
radiative envelopes, their disk accretion mechanism would be
different to that for the lower mass stars, where the magnetically
controlled accretion operates. In this context it may be useful to
note that earlier in its evolution, the star may be swollen due to the
large accretion ratess \citep{2010ApJ...721..478H} and generate a
magnetic field at this stage \citep{hoare2007}. Current evidence for
pre-Main Sequence Herbig Ae/Be stars suggests the transition
in disk accretion mechanism occurs around the A-B spectral type
boundary (\citealt{mottram2007, ababakr2016}). One of the few models
put forward to explain the accretion mechanism in massive Herbig Be
stars is the boundary layer where the circumstellar disk extends to
the surface of the central star. For example, through modelling the UV
spectra of a sample of Herbig Ae/Be stars, \citet{blondel2006}
suggested these intermediate to massive stars could accreting via a
boundary layer.\\
Next to the theoretical considerations, massive star formation poses a
considerable observational challenge. High-mass protostars are rare -
as predicted by the initial mass function described by
\citet{1955ApJ...121..161S} - and evolve fast, as the Kelvin-Helmholtz
timescale for contraction to a main sequence (MS) configuration is
shorter than the gravitational free-fall timescale. Unlike low-mass
stars, they are still embedded in their parental cloud whilst
accretion is ongoing. As such, they are both further away and suffer
from more dust extinction than their lower mass
counterparts. Therefore, observations at longer wavelengths are
essential for observations of high-mass star forming regions. An
important evolutionary stage for understanding the massive star
formation process is the Massive Young Stellar Object phase
(MYSO). These are objects which are still accreting material from
their birth cloud. They are radio-weak, having not yet ionized the
surrounding region to produce an H{\sc ii} region, in spite of their
high luminosities. MYSOs are frequently associated with jets and
bipolar outflows, likely driven by accretion activity. Information
about the accretion process can be obtained by studying these objects,
and then used to further refine theoretical models of massive star
formation. Due to dust extinction, they are only visible at infrared
wavelengths and longer. Overviews of the class of MYSOs can be found
in \citet{2013ApJS..208...11L} and \citet{2014EAS....69..319O}.\\
Near-infrared (NIR) techniques can be particularly useful to study the
inner 100s of au, where an accretion disk is expected to be
found. Using interferometry at 2$\mu$m, \citet{2010Natur.466..339K}
directly imaged a $\sim$15 au scale disk around the 20
M$_{\odot}$ MYSO G310.0135+00.389, lending support to disk
accretion. However, due to the limited sensitivity of current
instruments, this technique can only be applied to the brightest
targets, and this result has so far been the only such
detection. Another approach to studying accretion disks is to look for
kinematical tracers among observed NIR spectral transitions. The CO
ro-vibrational lines ("bandheads") at 2.3$\mu$m have been modelled by
\citet{1995ApJ...446..793C}, \citet {2010MNRAS.408.1840W} and
\citet{ilee2013} to arise in warm (4000K) and dense
(10$^{10}$cm$^{-3}$) self-shielded neutral disk material in
MYSOs. However, high signal-to-noise (SNR) observations at high
spectral resolution are required in order to model the bandheads (as
\citealt{ilee2013} note), which requires a large observing program for statistically significant conclusions about the presence of
disks in MYSOs to be inferred. In addition, only a small fraction of
these objects show these transitions (17 \% detection rate by
\citealt{2013MNRAS.430.1125C}, CLO13 from here).\\
Stellar winds are traced by hydrogen recombination lines, particularly
the Brackett series in the NIR, as modelled by
\citet{1981ApJ...251..552S}. The hydrogen recombination profiles
provide clues about the line emitting region. This is possible due to
the differences in the optical thickness of the various lines.\\
By considering the ratio of two hydrogen recombination lines,
\citet{1993MNRAS.265...12D} have shown that for S106 IRS1, the
hydrogen emission originates from a fast (100s km/s), optically thick
wind combined with a, slower (10s km/s), optically thin nebular
component. \citet{1995MNRAS.272..346B} extended this analysis to 6
more objects. They found that some objects show similar profiles to
S106 IRS1, whilst others display behaviour suggestive of an
accelerated stellar wind. They interpret differences in the optically
thin component of the line ratio as a sign of evolutionary
effects. The narrow peak will be stronger in MYSOs at later
evolutionary stages, when the increasing ionizing flux generates an
optically thin nebular component.
Thus far, the NIR spectrum of MYSOs has been studied either in small
samples at high spectral resolution, but small wavelength coverage
(e.g. \citealt{ilee2013} observed a dozen objects at $R \sim 30,000$
across CO 2.3 $\mu$m; \citealt{blum04} observed 4 MYSOs at $R \sim 50,000$ across the same region; \citealt{bik06} studied 20 MYSOs in the K band (between 2.08-2.18 and 2.28-2.40 $\mu$m) at $R \sim 10,000$;) or in large samples at broader wavelength coverage (e.g. CLO13 observed around 250 objects at $R \sim 500$ in $H$ and $K$). In this paper we present medium resolution
(R$ \sim$ 7000) near-infrared spectra of 36 MYSOs selected from the RMS
catalogue presented by \citet{2013ApJS..208...11L}.  This is the
largest sample of MYSOs studied in this manner to date. The resolution
allows studies of line profiles and detection of fainter features when
compared to the lower resolution observations while the larger
wavelength coverage allows for the study of a range of lines with
varying excitation properties.\\
Section \ref{sec:obs} describes the sample, observations, and data
reduction. In Section \ref{sec:av} we present our assessment of dust
extinction. Section \ref{sec:results} contains the main analysis of
this paper; MYSO spectral type measurements, correlations
between different line luminosities and features of the H
recombination line profiles. In section \ref{sec:concl} we present the
conclusions of this research.

\section{Observations}
\label{sec:obs}

The data were obtained between 18 February and 23 July 2011 using the
NIR spectrograph GNIRS on the Gemini-North telescope. 36 objects were
observed over 20 nights. We used the cross-dispersed mode, coupled
with the 'short' 111 l/mm camera, and a 2 pixel wide slit. Each pixel
is 0.15", giving a slit width of 0.3" and length of 7". The average
seeing was 0.75". Since the objects are bright at $K$, the data were
taken as a bad-weather programme, the conditions were not photometric.
The observed wavelength ranges are set up in orders and centered
around the positions of the {\it K, H, J} and {\it X} photometric
bands. The $K$ band ranges between 2.1472 and 2.3355 $\mu$m, the $H$
band between 1.6107 and 1.7519 $\mu$m, the $J$ band between 1.2888 and
1.4017 $\mu$m and the X band between 1.0741 and 1.1682 $\mu$m. In the
following, all wavelengths for spectral lines are in vacuum.  Because
the sources are red, the photon count and thus the SNR is highest at
longer wavelengths.
  
\begin{landscape}

    \begin{table}
    
\caption{Source list, including date, exposure time, position angle of
  the slit and known properties.  1 -Magnitudes from 2MASS; 2 - extinctions from $H$ band continuum slope
  derived in this work; 3 - LSR velocities from
  \citet{2011MNRAS.418.1689U}; distances from
  \citet{2011MNRAS.410.1237U}; bolometric luminosities calculated by
  \citet{2011A&A...525A.149M} * - sources not in RMS database, distances and luminosities
  from \citet{1981ApJ...251..552S}. ** - As the binarity of G110 is a new discovery and the slit was oriented E-W, we cannot accurately pinpoint the declination of the two objects. We believe the distance between them to be no more than 0.75" as given by the seeing. Luminosities, distances and LSR velocities for visual binaries given as for the whole system.}

\centering
\begin{tabular}{llrllrllllrrrr}
\cline{1-14}
Date   & RMS Name              & Integration & RA         & Dec.        & PA  & m$_{J}$$^{1}$ & m$_{H}$ $^{1}$ & m$_{K}$$^{1}$ & A$_{V}$             & V$_{\rm LSR}$                        & Distance     & L$_{bol}$ & Other       \\
(2011) &                       & time (s)    & (J2000)    & (J2000)     &     &               &               &               &       (mag)$^{2}$               & $(\rm km/s)^{3}$ & (kpc) $^{4}$ &   (L$_{\odot}$)$^{5}$                           & name        \\ \cline{1-14}
20.04  & G010.8411-02.5919     & 960         & 18:19:12.1 & -20:47:30.9 & 90  & 16.3          & 13.2          & 9.7           & 55                   $\pm$  3  & 12.3                             & 1.9          & 24000                        &   GGD 27          \\
29.06  & G010.8856+00.1221     & 960         & 18:09:08.0 & -19:27:24.0 & 90  & 15.6          & 12.9          & 9.6           & 51                   $\pm$  4  & 19.7                             & 2.7          & 5500                         &             \\
16.05  & G012.9090-00.2607     & 960         & 18:14:39.6 & -17:52:02.3 & 60  & 15.3          & 13.2          & 9.2           & 30                   $\pm$  3  & 36.7                             & 2.4          & 32000                        & W33A        \\
16.05  & G014.9958-00.6732     & 64          & 18:20:19.5 & -16:13:29.8 & 90  & 12.7          & 9.8           & 7.3           & 39                   $\pm$  14 & 19.4                             & 2            & 13000                        & M17SW IRS1  \\
29.06  & G015.1288-00.6717     & 480         & 18:20:34.6 & -16:06:28.2 & 90  & 12.9          & 10.5          & 8.9           & 25                   $\pm$  2  & 19.0                             & 2            & 12000                        &             \\
29.05  & G017.6380+00.1566     & 64          & 18:22:26.4 & -13:30:12.0 & 90  & 15.0          & 12.7          & 7.3           & 93                   $\pm$  3  & 22.1                             & 2.2          & 100000                       &  AFGL 2136           \\
18.06  & G018.3412+01.7681     & 960         & 18:17:58.1 & -12:07:24.8 & 0   & 15.3          & 12.9          & 9.3           & 53                   $\pm$  7  & 33.1                             & 2.8          & 22000                        &             \\
9.07   & G023.3891+00.1851     & 304         & 18:33:14.3 & -08:23:57.4 & 90  & 15.8          & 11.6          & 8.4           & 47                   $\pm$  3  & 75.5                             & 4.5          & 24000                        &             \\
16.07  & G025.4118+00.1052\_A    & 480         & 18:37:16.9 & -06:38:29.8 & 90  & 17.2          & 15.7          & 12.9          & 61                   $\pm$  1  & 95.3                             & 5.2          & 9700                         &             \\
23.07  & G026.2020+00.2262     & 456         & 18:38:18.5 & -05:52:57.4 & 90  & 14.2          & 10.6          & 8.3           & 37                   $\pm$  4  & 112.4                            & 7.5          & 3600                         &             \\
20.07  & G026.3819+01.4057\_A    & 480         & 18:34:25.7 & -05:10:50.2 & 90  & 13.1          & 10.8          & 9.1           & 33                   $\pm$  6  & 42.1                             & 2.9          & 17000                        &             \\
22.07  & G027.7571+00.0500     & 960         & 18:41:48.0 & -04:34:52.9 & 90  & 16.7          & 13.0          & 9.3           & 59                   $\pm$  16 & 99.6                             & 5.4          & 13000                        &             \\
29.06  & G029.8620-00.0444     & 960         & 18:45:59.6 & -02:45:06.5 & 50  & 15.1          & 12.5          & 9.8           & 43                   $\pm$  5  & 101.2                            & 4.9          & 28000                        &             \\
20.05  & G030.1981-00.1691     & 960         & 18:47:03.1 & -02:30:36.1 & 90  & 17.0          & 12.6          & 9.3           & 53                   $\pm$  4  & 103.1                            & 4.9          & 30000                        &             \\
20.05  & G033.3891+00.1989     & 64          & 18:51:33.8 & +00:29:51.0 & 90  & 13.0          & 9.6           & 7.2           & 33                   $\pm$  4  & 85.3                             & 5            & 13000                        &    GGD 30 IRS 3         \\
20.07  & G033.5237+00.0198     & 960         & 18:52:26.7 & +00:32:08.9 & 90  & 15.8          & 12.0          & 8.9           & 45                   $\pm$  4  & 103.5                            & 7            & 13000                        &             \\
22.07  & G034.0500-00.2977\_A  & 304         & 18:54:31.9 & +00:51:32.6 & 90  & 14.1          & 12.2          & 11.0           & 11                   $\pm$  3  & 11.5                             & 12.9         & 23000                        &             \\
22.07  & G034.0500-00.2977\_B  & 304         & 18:54:32.3 & +00:51:33.2 & 90  & 11.0          & 9.6          & 8.4           & 21                   $\pm$  4  &                              &          &                         &             \\
23.07  & G034.7123-00.5946     & 960         & 18:56:48.3 & +01:18:47.1 & 90  & 18.4          & 13.4          & 9.2           & 59                   $\pm$  4  & 44.5                             & 2.9          & 9700                         &             \\
22.04  & G056.4120-00.0277     & 160         & 19:36:21.5 & +20:45:17.9 & 90  & 12.5          & 9.9           & 8.1           & 31                   $\pm$  2  & -4.4                             & 9.3          & 22000                        &             \\
10.06  & G073.6525+00.1944     & 960         & 20:16:22.0 & +35:36:06.2 & 55  & 13.9          & 11.5          & 9.6           & 31                   $\pm$  7  & -73.4                            & 11.2         & 100000                       &             \\
25.06  & G073.6952-00.9996     & 120         & 20:21:18.9 & +34:57:50.9 & 0   & 12.9          & 10.6          & 8.0           & 47                   $\pm$  5  & -31.3                            & 7.4          & 17000                        &             \\
16.07  & G076.3829-00.6210     & 24          & 20:27:26.8 & +37:22:47.7 & 90  & 10.4          & 7.7           & 5.9           & 31                   $\pm$  5  & -1.7                             & 1.4          & 40000                        & S106 IRS1   \\
10.06  & G077.4622+01.7600\_A    & 480         & 20:20:39.3 & +39:37:58.5 & 150 & 12.8          & 10.6          & 8.9           & 39                   $\pm$  4  & 2.1                              & 1.4          & 3100                         &             \\
24.06  & G078.8867+00.7087     & 40          & 20:29:24.0 & +40:11:19.4 & 0   & 14.3          & 10.8          & 6.6           & 65                   $\pm$  2  & -6                               & 3.3          & 200000                       & AFGL 2591   \\
23.06  & G094.3228-00.1671     & 960         & 21:31:45.1 & +51:15:35.3 & 90  & 15.4          & 12.0          & 9.8           & 47                   $\pm$  18 & -38.4                            & 4.4          & 5700                         &    CPM 15         \\
16.07  & G094.6028-01.7966     & 48          & 21:39:58.3 & +50:14:20.9 & 90  & 10.9          & 9.2           & 6.8           & 29                   $\pm$  3  & -43.9                            & 4.9          & 43000                        & V645 Cygni  \\
21.07  & G102.3533+03.6360     & 48          & 21:57:25.2 & +59:21:56.6 & 90  & 12.3          & 9.6           & 7.2           & 39                   $\pm$  2  & -88.6                            & 8.4          & 110000                       &    CPM 36         \\
22.07  & G106.7968+05.3121*    & 24          & 22:19:18.2 & +63:18:47.0 & 64  &               & 8.1           & 6.1           & 59                   $\pm$  1  &                                  & 0.9          & 13000                        & S140 IRS1   \\
21.07  & G110.1082+00.0473B\_A  & 480         & 23:05:10.2 & +60:14:42.7** & 90  & 10.7          & 10.0          & 10.1          & 3                    $\pm$  3  & -52.1                            & 4.3          & 17000                        &             \\
21.07  & G110.1082+00.0473B\_B  & 480         & 23:05:10.3& +60:14:42.7** & 90  &           &   9.8        &   9.5        & 3                    $\pm$  1  &                             &          &                         &             \\
26.06  & G111.2348-01.2385     & 960         & 23:17:21.0 & +59:28:48.0 & 10  & 14.1          & 11.3          & 9.6           & 73                   $\pm$  3  & -54.4                            & 4.4          & 42000                        &    IRAS 23151+5912         \\
21.07  & G111.5234+00.8004A    & 96          & 23:13:32.4 & +61:29:06.2 & 45  & 11.1          & 9.7           & 7.8           & 35                   $\pm$  3  & -58.6                            & 2.6          & 5600                         &     NGC 7538 IRS 4        \\
25.06  & G111.5423+00.7776     & 200         & 23:13:45.4 & +61:28:10.3 & 80  & 14.6          & 11.6          & 8.5           & 91                   $\pm$  3  & -57.2                            & 2.6          & 210000                       &             \\
21.07  & G120.1483+03.3745     & 48          & 00:23:57.0 & +66:05:51.5 & 90  & 11.8          & 9.0           & 7.0           & 33                   $\pm$  4  & -68.9                            & 5.6          & 21000                        &    CPM 1         \\
18.02  & G151.6120-00.4575     & 48          & 04:10:11.9 & +50:59:54.4 & 90  & 10.9          & 8.9           & 7.1           & 27                   $\pm$  3  & -49.7                            & 6.4          & 61000                        &     CPM 12        \\
18.02  & G213.7040-12.5971\_A* & 80          & 06:07:47.8 & -06:22:56.2 & 14  & 13.1          & 11.4          & 7.2           & 31                   $\pm$  2  &                                  & 0.95         & 25000                        & Mon R2 IRS3A \\
18.02  & G213.7040-12.5971\_B* & 80          & 06:07:47.9 & -06:22:55.4 & 14  &               &  9.5          &  7.3          & 31                   $\pm$  2  &                                  &          &                         & Mon R2 IRS3B \\
18.02  & G233.8306-00.1803     & 24          & 07:30:16.7 & -18:35:49.1 & 60  & 10.9          & 8.0           & 6.1           & 35                   $\pm$  14 & 44.6                             & 3.3          & 13000                        &             \\ \cline{1-14}
\label{tab:sources}
\end{tabular}
\end{table}
\end{landscape}

The observed sources were selected from the RMS database
\citep{2013ApJS..208...11L}. Two objects (G106.7968 and G213.7040) are
not in the RMS database as they are located at a high galactic
latitude. For these, the distance and bolometric luminosity are taken
from \citet{1981ApJ...251..552S}. The selection criteria included a
source classification as MYSO, observability from Gemini-N, and being
bright enough at $K$ to result in good SNR spectra at these
wavelengths. Emphasis was put on obtaining data from the most luminous
objects fulfilling those criteria. 90\% of the observed objects are
bright (L$>$8000 L$_{\odot}$) and radio-quiet, and all have L$>$3000 L$_{\odot}$ and radio flux$<$0.5 Jy at 5 GHz. As such, they likely have not yet started
to ionize the ISM and produce an HII region.  The sample properties
are described in the Table \ref{tab:sources}.

Most targets are compact enough to allow nodding along the slit in an
ABBA sequence to remove the sky background. For extended sources a sky
region outside of the nebula was used. Spectra of nearby main sequence
stars of types B9V-A4V were obtained in order to correct for
atmospheric telluric absorption. The airmass difference between these
and the science targets was always less than 0.1. Stars with these
spectral types were used as they have few intrinsic absorption
features in the wavelengths of interest to this work. Where necessary,
their absorption features were fitted and removed prior to telluric
line correction. A standard reduction procedure was applied to all the
spectra, using the \textit{PyRAF} package
(\citealt{2012ascl.soft07011S}). The bias and dark subtraction are
done in the pre-reduction phase at the telescope. Pixel-to-pixel
variations were corrected for by using flat-field frames. The sky
background was removed with the standard ABBA
subtraction. One-dimensional spectra from each order were then
extracted.  Argon lamp spectra and telluric absorption lines from the
solar spectrum of \citet{1995ASPC...81...66H} were used to perform the
wavelength calibration. The resulting calibration was assessed by
measuring the wavelengths of telluric lines. The wavelength
calibration was found to be accurate to $\sim$5 km s$^{-1}$.  The
resolution of the spectra as measured from the arc calibration lines
is in the range 35-49 km s$^{-1}$, corresponding to
R$\approx$7000. The $K$ and $H$ band spectra have high SNR ($>$100 for
most spectra), while 66\% of the $J$ and 50\% of the X bands have
SNR$>$30.

The intrinsic absorption features of the telluric standards were
removed by using Voigt profiles. This proved particularly difficult
for the Br10 line at 1.737 $\mu$m, as a number of telluric lines are
blended with this HI line. As such this recombination line was not
used in our analysis. The calibrated target spectra were then divided
by the telluric standard (with the intrinsic features removed), in
order to remove the atmospheric absorption features. The result is
multiplied by a blackbody of the temperature of the telluric standard
star to retrieve the relative photometric shape of the target
spectra. The spectra were not obtained in photometric conditions, as
accurate flux calibration was not the main aim of the
observations. However, when using differential count ratios between
the different bands in spectra of individual objects, the resulting
colour agreed to within 0.5 magnitude with those from 2MASS.

The fact that relative count rates return fairly good estimates of the
colours, allowed us to determine the continuum fluxes of MYSOs in the
X band (1.09 $\mu$m). The scaling factor was derived by using the
relative count rates between X, $J, H,$ and $K$, with the respective
colours of the telluric standard stars (whose continuum flux at 1.09
$\mu$m were obtained by interpolating between their fluxes at the $J$
and $I$ photometric bands). The MYSO spectra were then scaled at the
given 2MASS central band wavelengths.

For the computation of line fluxes, we use catalogued photometry from
2MASS \citep{2006AJ....131.1163S}, We also considered UKIDSS
\citep{2008MNRAS.391..136L}, but find that using the UKIDSS photometry
results in an discontinuous spectrum between different orders for a
large fraction of our sample.  The reason for this is the sensitivity
of the UKIDSS survey. Many of the targets have m$_{K}<$10, which often
leads to saturation in UKIDSS. Where 2MASS data was not available, the MYSO magnitudes were estimated from the
telluric star magnitude, with as caveat that the fluxes carry large
errorbars - 0.4 magnitudes in the K band.

The velocity from the Br$\gamma$ line was compared to values from the
RMS database, as reported by \citet{2011MNRAS.418.1689U}. The deduced
LSR velocities agree very well with the RMS LSR velocities from
NH$_{3}$ surveys, with a Pearson correlation factor of 0.89, and a
scatter of 6.5 km/s around the best fitting line. The other
recombination lines yielded similar results, implying that the objects
are associated with the high density clouds probed by RMS.

\section{Extinction}
\label{sec:av}

The determination of the dust extinction to the target objects, and
the subsequent correction for it, are problematic in the case of
MYSOs.  As explained in the introduction, these regions are
considerably affected by dust. In the sample of CLO13 of 135 MYSOs,
the extinctions found ranged between 2.7$<$A$_{V}<$114. \\
We estimate the extinction using the slope of the continuum, in a
similar manner to \citet{1998A&A...332..999P}. The spectra of MYSOs
show a rising red continuum, which is caused by a combination of
reddening of the stellar photosphere by dust extinction on the one
hand and thermal emission by dust on the other hand.
\citet{1990ApJ...357..113M} derive that
$A_{\lambda}\propto\lambda^{-1.8}$ between 1-4$\mu$m (see also
\citealt{moore2005} for the case of high $A_V$). The Rayleigh-Jeans
(RJ) approximation can be used for a blackbody at NIR wavelengths,
i.e. $F_{\lambda}\propto\lambda^{-4}$. We determine the value of the
extinction that produces a dereddened spectrum with a slope closest to
that of the RJ approximation through chi-squared minimisation. We
compute values for A$_{V}$ for each of the spectral bands of an object
in this manner. The errors are determined from the chi-squared
analysis. The $K$ band yields the highest value for A$_{V}$, this is
most likely due to dust emission starting to contribute significantly
at these wavelengths.  \citet{1998A&A...332..999P} demonstrated that
extinctions derived from the continuum slope at the shorter $J, H$
bands are in agreement with those from the line fluxes of the hydrogen
lines assuming case B recombination\footnote{Their observed hydrogen
  recombination lines covered a larger wavelength range than ours,
  which prevents us from making a similar estimate.}. Most of our sources are too embedded to be detected in the J or X bands, or have too low an SNR for extinction estimations from the continuum slope. We therefore choose to use the extinction obtained from the spectral slope of the $H$ band continuum in all of the following analysis, as the effect of hot dust excess is considerably lower than in the $K$ band. The values are listed in Table \ref{tab:sources}.

\begin{figure*}
\centering
\includegraphics[scale=0.33]{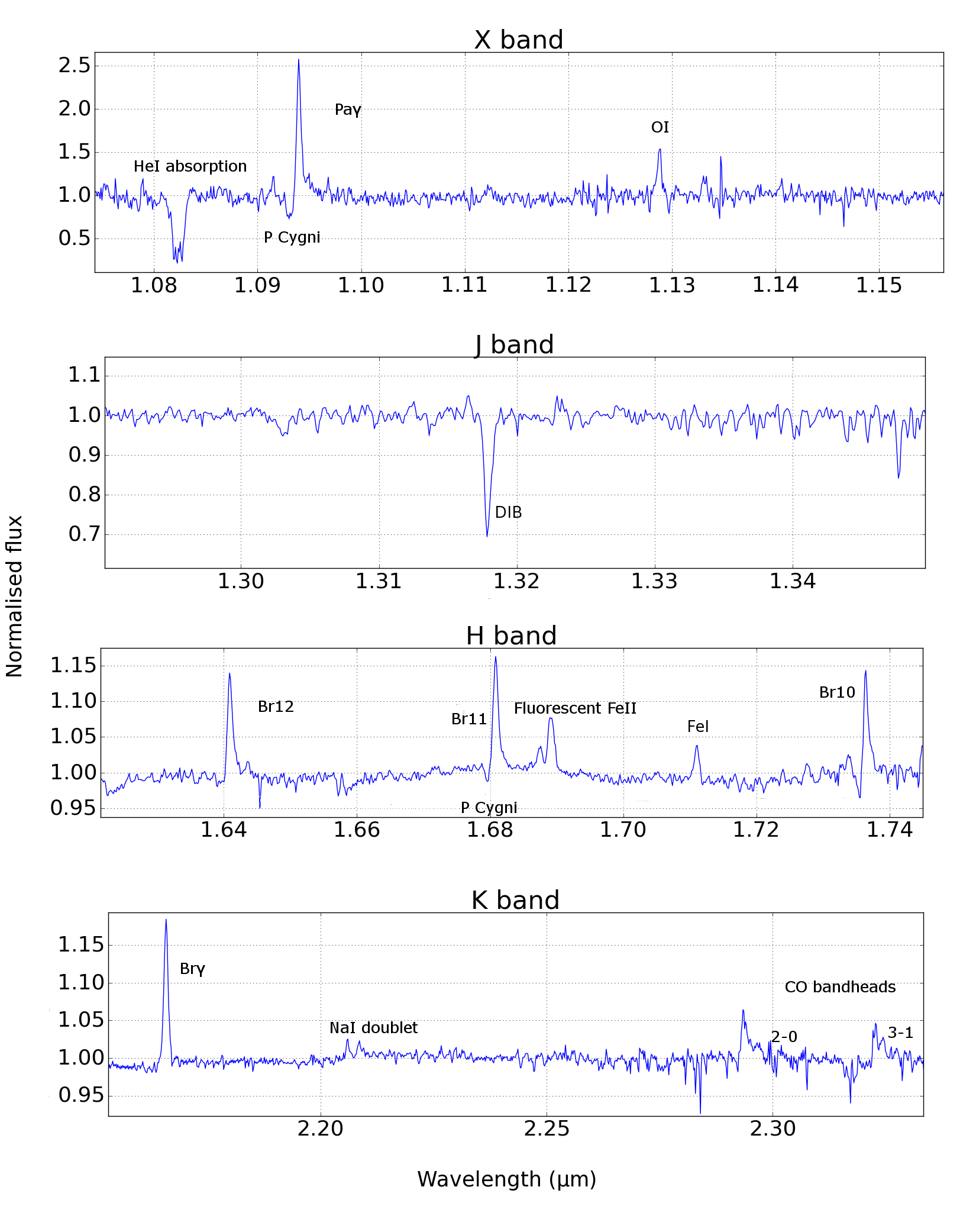} 
\caption{Example spectrum of G056.4120-00.0277, continuum normalized. From bottom to top, the data correspond to the \textit{K}, \textit{H}, \textit{J} and \textit{X} bands respectively.}
\label{fig:g056}
\end{figure*}

\section{Results}
\label{sec:results}
We begin this section with an overview of the spectra and their
characteristics. We then discuss the absorption lines focussing on the
hydrogen lines that are found in absorption towards 2 objects,
allowing their spectral types to be derived. This is followed by an
analysis of the emission lines, their fluxes and their profiles, while
hydrogen emission line profile ratios are used to learn about the line
formation regions.

\begin{figure*}
\centering
\includegraphics[scale=0.22]{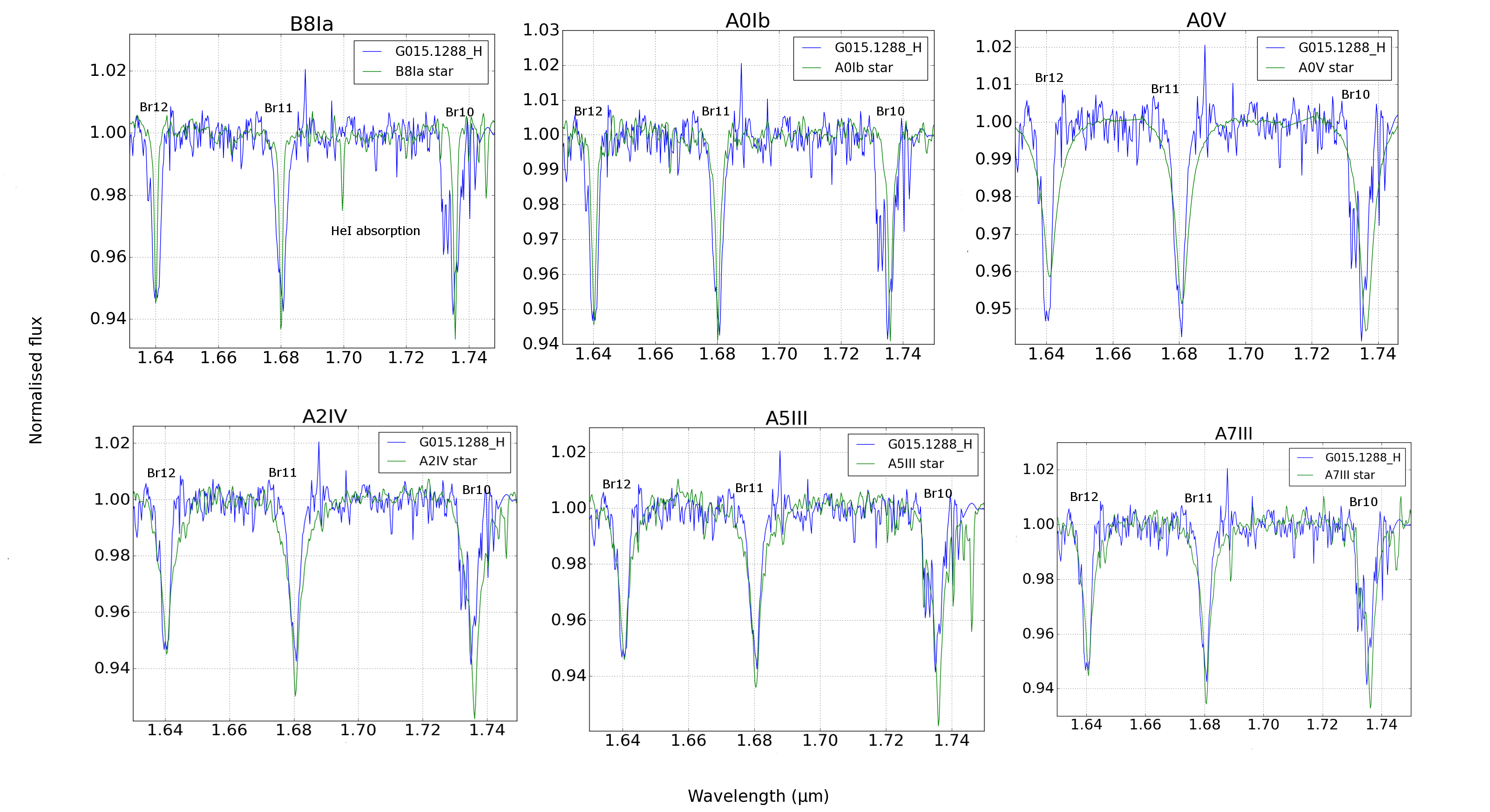} 
\caption{Br10-12 profiles in G015.1288-00.6717 (blue) compared with normal stars with added continuum dust excess (green). Data from \citet{1998ApJ...508..397M} and \citet{2005ApJS..161..154H}. See text for details.} 
\label{fig:dust_g015spt}
\end{figure*}

\subsection{Description of spectra}

An example spectrum, of G056.4120-00.0277, is presented in Figure
\ref{fig:g056}.  The HeI transition at 1.083$\mu$m, as well as the
Pa$\gamma$ hydrogen recombination line at 1.094 $\mu$m are present in
the X-band. Strong telluric absorption hinders the detection of
spectral lines in the red part of this band. Strong telluric
absorption minimizes the use of the $J$ band past $\sim$1.35$\mu$m.
The only transitions in the observable range are the OI line at
1.316$\mu$m, which is seen towards some sources, as well as strong
Diffuse Interstellar Band (DIB) absorption at 1.318$\mu$m. The $H$
band spectrum is marked by three strong Brackett series transitions -
12-4 at 1.641 $\mu$m, 11-4 at 1.681 $\mu$m and 10-4 at 1.737$\mu$m. A
number of Fe transitions are also present, such as the shocked [FeII]
transition at 1.644 $\mu$m, the fluorescent FeII transition at 1.689
$\mu$m, as well as another FeI line at 1.711 $\mu$m.  The strongest
feature in the $K$ band is the Br$\gamma$ line at 2.166 $\mu$m. Other
transitions observed in the $K$ band are the NaI doublet at 2.206 and
2.209 $\mu$m, and the shocked H$_{2}$ transition at 2.24
$\mu$m. Finally, the first two CO first-overtone bandhead transitions,
located at 2.29 and 2.32 $\mu$m are also covered.

\begin{table*}
  \centering
  \caption{Detection rates of various features observed in our sample,
    compared to their respective values in the CLO13 survey. A number
    of lines that we find were not detected in the previous
    survey. The higher spectral resolution allows for fainter lines to
    be detected, while the current $H$ wavelength coverage is broader
    than in CLO13.}
\label{tab:detrates}
\begin{tabular}{llllllllllllll}
\hline
Spectral line        & Br$\gamma$ & Br12 & Br11 & Br10 & Pa$\gamma$ & H2   & NaI & CO emission  & [FeII] & fluorescent FeII & OI   & HeI  \\
\hline
Detection rate      & 97\%       & 82\% & 79\% & 79\% & 26\%       & 21\% & 37\%        & 34\%           & 42\%  & 61\%             & 21\% & 24\% \\
CLO13 rate & 75\%       & 44\% & 37\% & 45\% & N/A        & 9\%  & N/A         & 17\%    & 34\%           & 26\%             & N/A  & N/A   \\
\hline
\\
\end{tabular}
\end{table*}

The detection rates of various lines are presented in Table
\ref{tab:detrates}. Virtually all objects have Br$\gamma$ in
emission. The detection rate of the higher Brackett lines is slightly
smaller, which can be attributed to the fact that these lines are
intrinsically weaker, and therefore less prone to detection. Overall,
the detection rates are much higher than in CL013, which can be
readily explained by the superior spectral resolution employed here,
facilitating the detection of weaker lines that were unresolved by
CLO13.\\
Three objects (G034.0500-00.2977, G110.1082+00.0473B and
G213.7040-12.5971) showed two separate spectra within the slit, which,
when extracted, appear to be objects with different spectra, and which
thus may form a binary system. G034.0500 is referred to in the RMS
database as a binary source, whereas the binarity of G213.7040 and G110.1082 are new findings. However, in the absence of other data to confirm binarity we stress that it is possible these are visual binaries rather than bound systems.

\subsection{Absorption lines}

\begin{figure}
\centering
\includegraphics[scale=0.31]{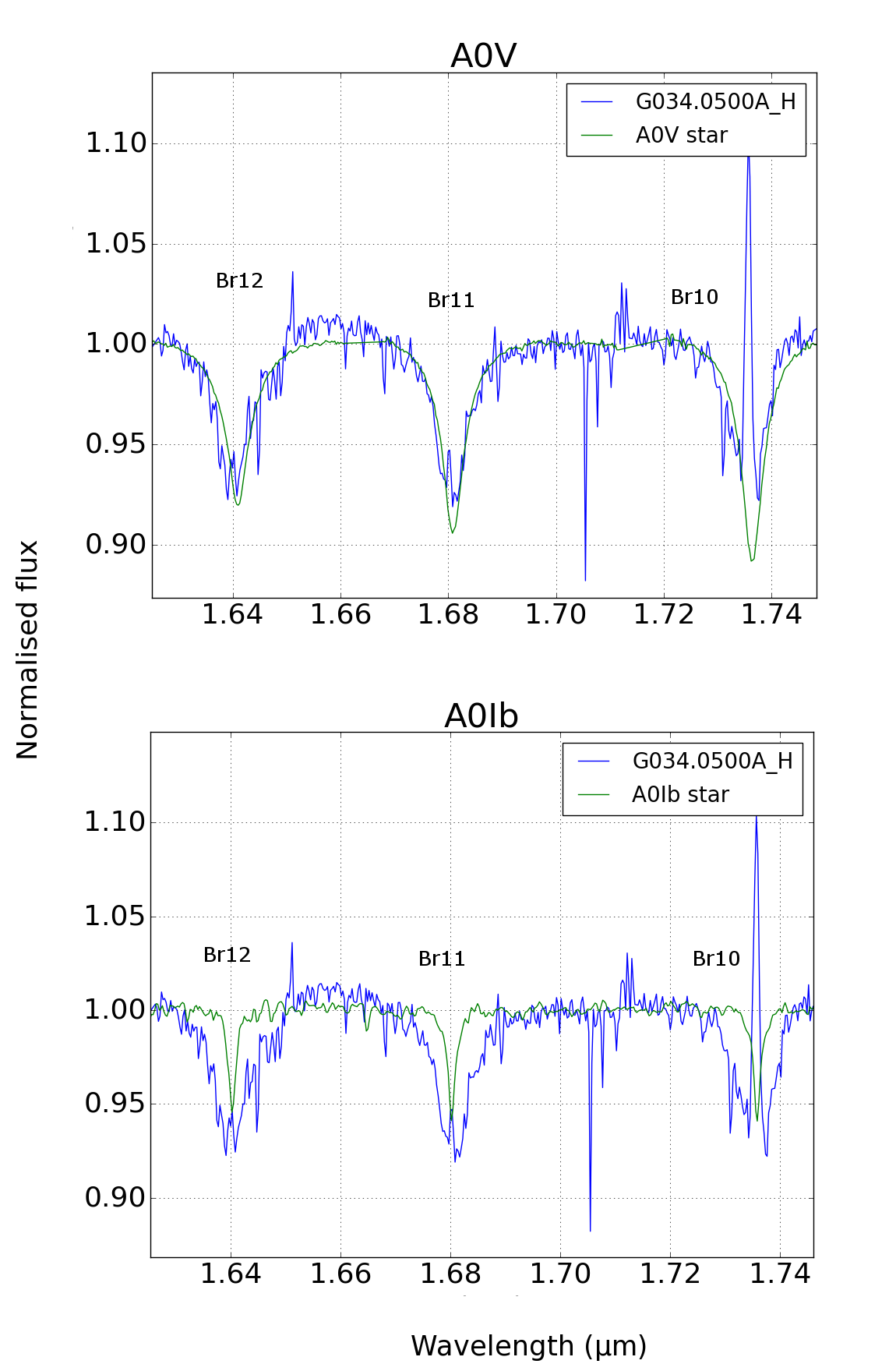} 
\caption{Spectral typing of G034.0500\_A. The Brackett absorption lines are too large compared to what would be expected of a A0Ib star, they are more like an A0V.}
\label{fig:g034spt}
\end{figure}

\subsubsection{Spectral typing}

The stellar parameters can not be determined in the usual manner as
the targets are invisible at blue wavelengths - the classical wavelength
regime for spectral typing.  At near-infrared wavelengths the MYSO
photospheres are heavily veiled by dust emission, and spectral typing is
difficult as well. In practice, spectral types of MYSOs are usually
estimated using the total luminosity derived from Spectral Energy
Distribution (SED) fitting.\\
Traditionally, MYSOs have been found to have luminosities similar to
those of OB stars.  Spectral types have rarely been determined
directly in MYSOs, and not at all in the case of RMS MYSOs.
Previously, \citet{2003A&A...408..313K} observed $H$-band spectra of
three MYSOs, IRAS 17175-3544, 17441-2910 and 18079-1756 at
intermediate resolution (R$\sim$5000). They found photospheric
absorption lines in IRAS 18079-1756 and IRAS 17175-3544, and assigned
spectral types of B3V and O7-8V respectively to these two objects.

Most of our MYSOs show strong, broad HI emission, due to a stellar
wind (see also Sec. \ref{sec:emlines}). However, two of the objects in
this sample (G015.1288-00.6717 and G034.0500-00.2977\_A) display
absorption in the Br10, Br11 and Br12 lines. There are also hints of
absorption in Br$\gamma$ and Pa$\gamma$, alongside the dominant
emission in these two objects. The RMS survey quotes G015.1288 as
having L$_{bol}$=13000 L$_{\odot}$. This corresponds to a spectral
type of B0.5 for a Zero-Age Main-Sequence star, using the values from
\citet{1981Ap&SS..80..353S}.

The absorption profiles in the $H$ band of G015.1288 were compared
with spectra of normal stars from the surveys of
\citet{2005ApJS..161..154H} and \citet{1998ApJ...508..397M}.  A
spectral type of F or later would be hard to reconcile with the MYSO
data, as F stars have a large number of metal lines. In addition, early O
spectral types do not match either, as their Brackett lines are weaker
and narrower than observed for G015. Late O and B1-5 stars show a HeI
line at 1.70 $\mu$m, at comparable or higher strength than the
hydrogen absorption lines, and is not present here. We note that the
near-infrared spectral surveys of \citet{2005ApJS..161..154H} and
\citet{1998ApJ...508..397M} do not have complete spectral and
luminosity type coverage. The only B type sample spectra available are
those for B1, B5 and B8.

\begin{figure}
\centering
\begin{tabular}{cc}
\includegraphics[scale=0.3]{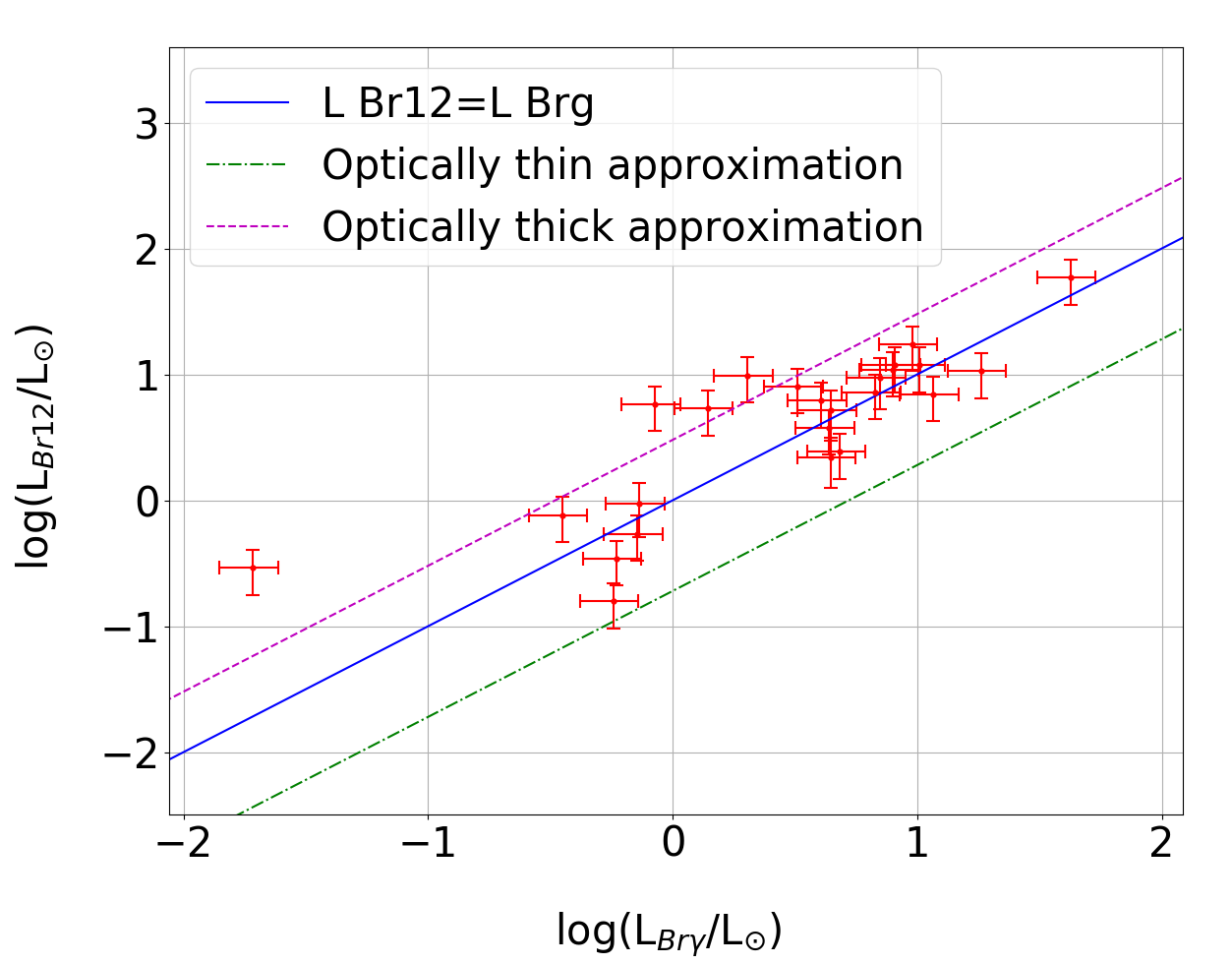} 
\end{tabular}
\caption{Br12 luminosity as a function of Br$\gamma$, with optically thin and optically thick approximations, and a equal luminosity line.}
\label{fig:br1210vg}
\end{figure}

So, given the presence and absence of diagnostic absorption lines, a
spectral type later than B5 and earlier than A9 would be expected for
G015.1288. However, the Br10-12 absorption in the B5-A9 stars is much
stronger than observed for G015.1288. It is possible that a
significant amount of excess could veil the HI absorption to a
comparable level to our observations. This is illustrated in Figure
\ref{fig:dust_g015spt}, which zooms in on the Brackett lines in the
$H$ band spectra of a number of normal B8-A7 stars with added excess
continuum emission. Overplotted on these are the GNIRS data for
G015.1288. A B8 star with added excess does not match the observed HI
profiles, as the lines in this star are considerably narrower than in
G015.1288. In addition, the HeI line is still strong enough to be
observed. The A0Ib profiles match the observations in strength, but
are narrower. An A0V star has broader wings in the Brackett lines than
G015.1288. In the absence of more observational Main Sequence star'
spectra to compare our data with, we conclude that G015.1288 has the
spectrum of an early A giant or supergiant, with an excess continuum
added to the stellar photosphere corresponding to 4$\times$the
continuum at $H$.

We applied a similar analysis to the Br10-12 absorption lines in the
spectrum of G034.0500-00.2977\_A. In this case the lines are
considerably broader, and we find the best fit is that of an A0V
(MS-like) star, as can be seen in Figure \ref{fig:g034spt}. In
contrast, the spectrum of G034.0500\_B shows the usual NIR rising red
continuum of MYSOs.\\
A full study of the binary system is beyond the scope of this
paper. However, given that a lower mass object will be in an earlier
evolutionary stage than a higher mass object of the same age, it is
more likely that G034.0500-00.2977\_A is a main sequence star located
in the same direction as source B, rather than a lower mass
star-forming object associated with a MYSO.

In summary, the (only) observed absorption spectrum of a MYSO in the
present sample can be reproduced by an A-type object with a relatively
low surface gravity. We need to invoke a continuum excess emission of
a factor of a few in order to match the lower depth of the lines. This
excess is due to radiation from hot dust, and does not affect the
spectral classification itself, as that is based on the linewidths and
relative strengths of aborption lines, both of which do not change
after the addition of a continuum excess. For a typical temperature of
10,000 K of an A0 star and an observed luminosity of 13,000
L$_{\odot}$, we infer a stellar radius of order 70 R$_{\odot}$ for G015.1288.

The fact that the star is cooler and larger than what we might expect
based on its total luminosity can be understood in
terms of the high accretion expected in massive star forming
objects. Constant accretion simulations done by
\citet{2010ApJ...721..478H}, as well as more recent hydrodynamic
simulations done by \citet{2016MNRAS.458.3299H} show that a pre-MS
star accreting at a constant rate of 10$^{-3}$ M$_{\odot}$yr$^{-1}$
can swell up to 40-300 R$_{\odot}$. With constant luminosity, this
puffing up will cause a decrease in temperature of a factor $\sim$3. A
puffed up late-O/early-B type star will have a temperature and
structure similar to an A-type giant or supergiant star. This might
explain why MYSOs have yet to ionize their surroundings and form an
HII region, as the UV flux of an A star is considerably lower than
that of a B star (cf. \citealt{hoare2007}).

\begin{figure}
\centering
\begin{tabular}{cc}
\includegraphics[scale=0.3]{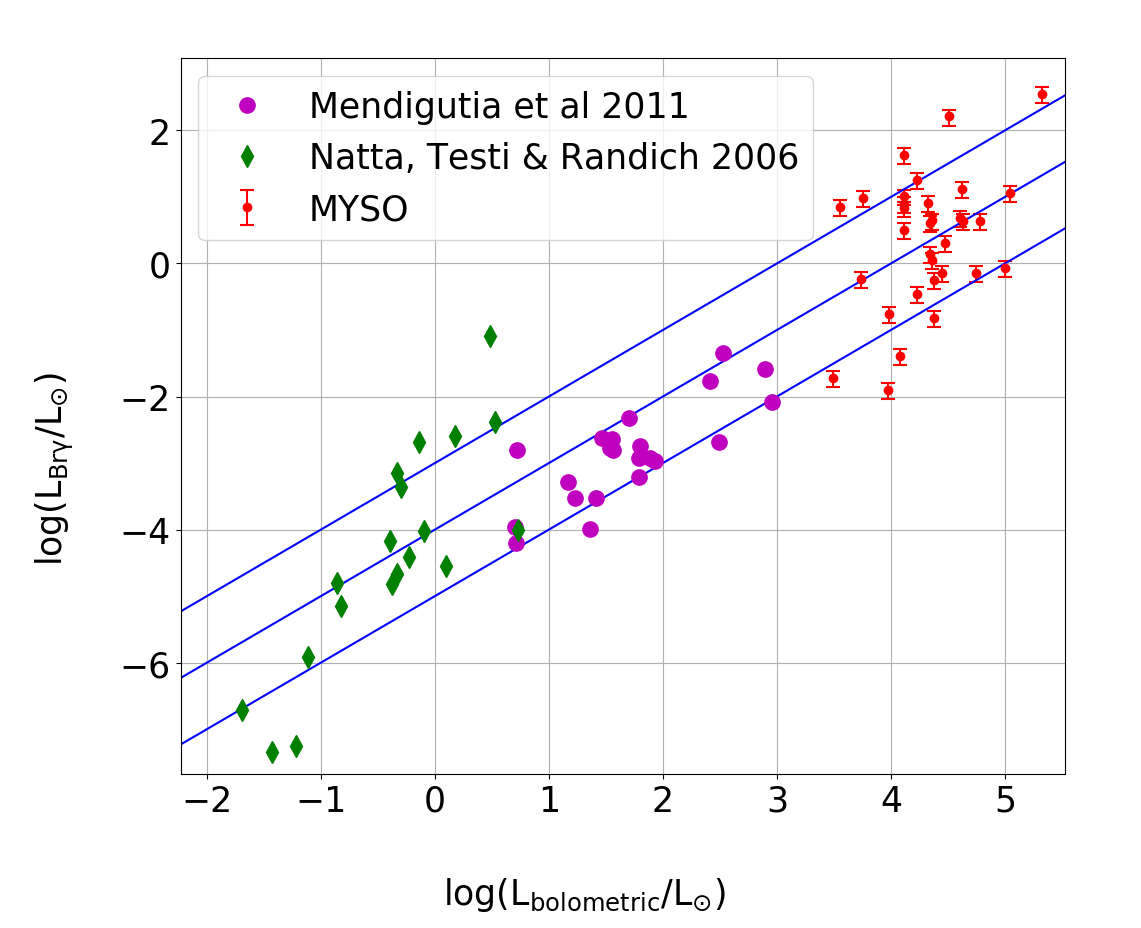} 
\end{tabular}
\caption{ Log-log plot of the luminosity in the Br$\gamma$ line
  against bolometric luminosity. Green diamonds are T Tauri stars,
  while purple circles are Herbig stars, objects of a lower and
  intermediate mass, respectively, believed to be at a similar
  evolutionary stage as MYSOs. The solid lines correspond to $\dot{\mathbf{L}}_{Br\gamma}$=10$^{-3}$L$_{bol}$, 10$^{-4}$L$_{bol}$ and 10$^{-5}$L$_{bol}$}
\label{fig:brgbol}
\end{figure}

\subsubsection{Other absorption lines}

Turning to other absorption lines, as can be seen in Figure
\ref{fig:g056}, the strongest absorption feature in the spectrum is
located at 1.318$\mu$m. We identify this absorption with the
$\lambda$13175 ($\rm \AA$) Diffuse Interstellar Band
(\citealp{cox2014,hamano2015}). It is detected in most objects for
which useable $J$-band data were obtained, and the equivalent widths
are measured to be up to several $\rm \AA$.  The lines in the MYSOs
are much stronger than reported thus far in the literature and this
can be partly understood due the fact that the extinction towards the
targets is much larger than those for example sampled in
\citet{cox2014} and \citet{hamano2015}, whose highest extinction lines
of sight have an $A_V$ of 10. These authors also determine the slope
of the relationship between the DIB's EW and $A_V$.  The EWs of the
band in the MYSOs do not follow these relationships however; the lines
are weaker than the $A_V$ values would imply. In understanding this
finding, we need to consider that the total extinction towards our
targets is a combination of foreground (interstellar) extinction and
extinction due to the dust in the parental molecular cloud and
circumstellar envelopes. It had been observed previously that the DIBs
do not trace circumstellar material as efficiently as they trace the
interstellar material. This could be due to different excitation
conditions in these respective environments, the net effect is that
the line-EWs are indeed lower than the total $A_V$ would suggest
\citep{oudmaijer1997}.

Finally, we also observe CO bandhead absorption lines, which is
expected to originate from a further out, colder component than its
emission counterpart \citep{2010MNRAS.402.1504D}. Studies of these
lines can help in determining the properties of the outer, cooler
envelope of MYSOs, but their analysis is beyond the scope of this
study.

\label{sec:spt}

\subsection{Emission lines}
\label{sec:emlines}

We have compiled an atlas of the spectral features found in the data
(presented in the Appendix). We consider a spectral line a detection
if it has a peak flux F$_{\lambda}$ $>$ 3$\delta$F$_{cont}$, and a
full-width half-maximum larger than the resolution limit where
$\delta$F$_{cont}$ is given by the root-mean-square variations of the
continuum counts.
The continuum normalised lines were fitted
with Gaussian profiles.

\subsubsection{Line luminosities and accretion rates}

Let us first discuss the line luminosities which are computed using
the distances and the measured Equivalent Width and the dereddened
continuum fluxes.

The lines in the Brackett series form in similar conditions. For
example, in the case B approximation, the ratio
Br$\gamma$/Br12~$\simeq$~5$\pm$1, for temperatures between 3000-30000K
and electron densities 10$^4$-10$^9$ cm$^{-3}$. In the optically thick
case, the Rayleigh-Jeans approximation applies and the ratio is given
by
Br$\gamma$/Br12~=~${(\lambda_{Br12}/\lambda_{Br\gamma})}^{-4}$S$_{Br\gamma}$/S$_{Br12}$,
where $S$ is the projected surface area the line is emitted from. If
both lines are emitted from an area of the same size, the ratio
becomes 0.33, the lower limit.  The Br12 and Br$\gamma$ luminosities
indeed correlate, as shown in Figure \ref{fig:br1210vg}. Most points
lie between the optically thin and optically thick approximations. The
optical thickness of these regions depends on the type of wind
present, which we will return to in Section \ref{sec:profiles}. CLO13
found that Br$\gamma$ luminosities lie between
log(L$_{Br\gamma}$)~=~log(L$_{bol}$)-3 and log(L$_{bol}$)-5. This
result is confirmed by the GNIRS data, as can be seen in Figure
\ref{fig:brgbol}. We also plot data of intermedaite mass pre-Main
Sequence Herbig Ae/Be stars from \citet{2011A&A...535A..99M} and low
mass T Tauri stars from \citet{2006A&A...452..245N} for comparison
with lower mass sources. These seem to also lie within the same ranges
as our data, indicating that there is some degree of continuity across
the mass range of different pre-Main Sequence objects.

\begin{figure}
\centering
\begin{tabular}{cc}
\includegraphics[scale=0.25]{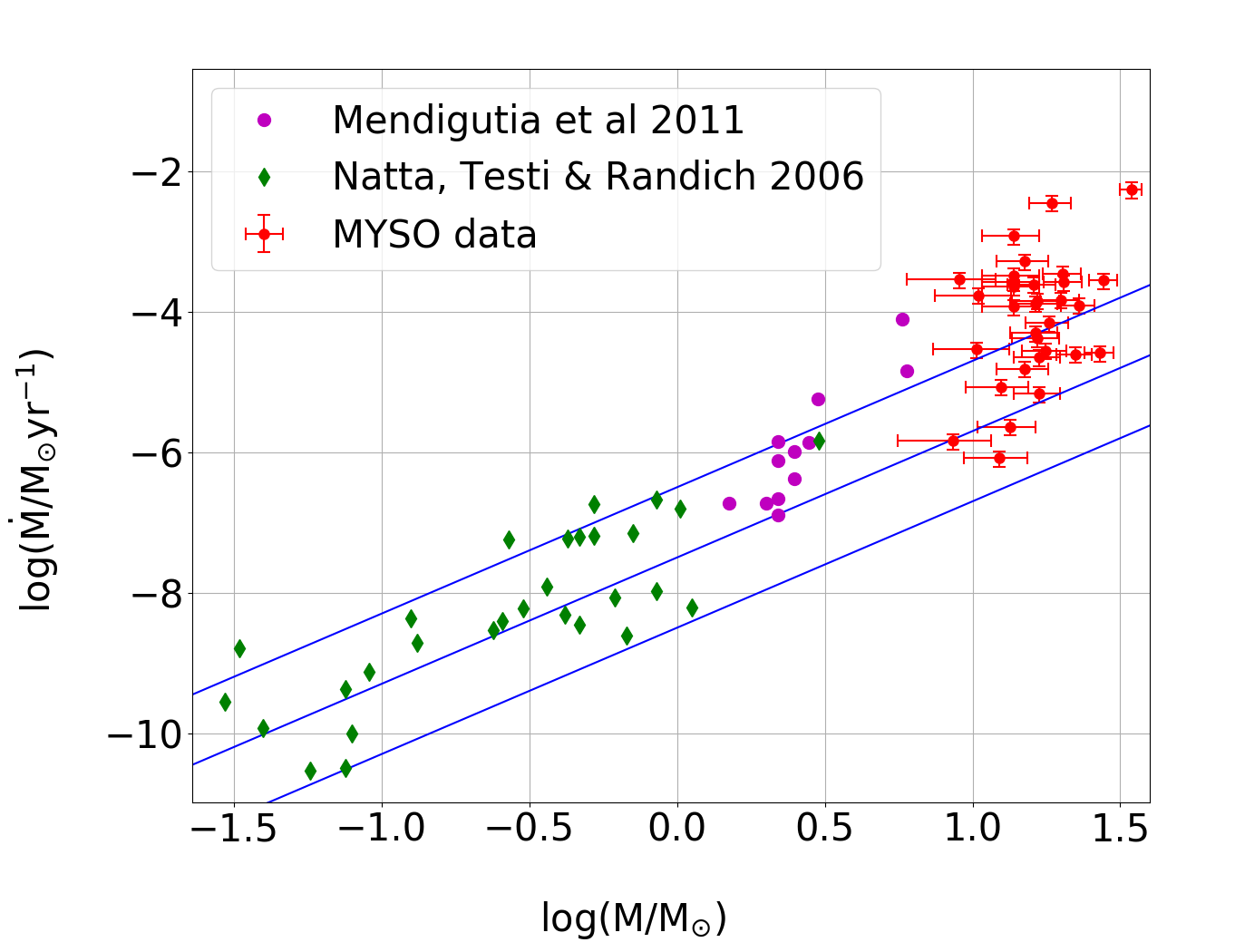}
\end{tabular}
\caption{ Log-log plot of the mass accretion rate against stellar mass. The symbols are as the previous figure. The solid lines correspond to $\dot{\mathbf{M}}$=10$^{-5.5}M_{*}$, 10$^{-6.5}M_{*}$ and 10$^{-7.5}M_{*}$}
\label{fig:accrates}
\end{figure}

The luminosity of the Br$\gamma$ line, and many other emission lines,
has been shown to be a proxy for the accretion luminosity in low and
intermediate-mass YSOs (\citealt{2011A&A...535A..99M};
\citealt{2006A&A...452..245N}). It stems from the observed correlation
between the Br$\gamma$ line luminosity with the accretion luminosity -
which is essentially the kinetic energy of the infalling material
converted into radiation in the accretion shock. The hydrogen
recombination line emission is seen to originate in the accretion
funnels in the lower mass T Tauri stars, whereas it appears to trace
winds in the higher mass Herbig Be stars (see
e.g. \citealt{fairlamb2015}). The reasons for the apparent
contradiction that the line luminosity - accretion luminosity
correlations hold in both situations can be intuitively explained by
the fact that the hydrogen is ionized by the radiation emerging from
the accretion shock region, leading to a relationship between
accretion and hydrogen recombination emission.  In principle,
therefore, we can exploit the correlation and proceed with computing
accretion luminosities from the Br$\gamma$ lines.

Following CLO13, we calculate accretion luminosities from
Br$\gamma$. To this end we use the \citet{2011A&A...535A..99M}
relation:

\begin{equation}
\log (L_{acc}/L_{\odot})=(3.55\pm0.80) + (0.91\pm0.27) \log(L_{Br\gamma}/L_{\odot})
\end{equation}

This results in accretion luminosities of MYSOs in the range
L$_{acc}$~=~0.1-10 L$_{bol}$. From the accretion luminosity we can
determine the mass accretion rate if we know the mass and radius of
the star, with the following formula:

\begin{equation}
\dot{M}_{acc}=\frac{L_{acc}R_{*}}{GM_{*}}
\end{equation}

Masses and radii are estimated from the bolometric luminosities using
the Zero Age Main Sequence relations as tabulated by
\citet{2011MNRAS.416..972D}, (M $\propto$ L$_{bol}^{0.33}$, R
$\propto$ L$_{bol}^{0.2}$).  Figure \ref{fig:accrates} shows the resulting
accretion rates against stellar mass. The mass accretion rates lie
above $\log(\dot{M}_{acc})=-6.5+1.8\log(M_{*})$ , similarly to lower mass
objects, but also extend to above the $\log(\dot{M}_{acc})=-8.5+1.8\log(M_{*})$ line, unlike the low mass sources.  It would appear that the results for the MYSOs follow the
trend of increasing mass accretion rate with mass as observed for the
lower mass T Tauri and Herbig Ae/Be stars, a trend that was also
pointed out by \citet{beltran2016} and CLO13. However, the MYSOs' mass
- mass accretion rate relation can be a fit by
$\log(\dot{M}_{acc})=(-7.0\pm1.4)+(2.4\pm1.2)\log(M_{*})$, which is a
steeper line than $\log(\dot{M}_{acc})=-7.5+1.8\log(M_{*})$, which was
the best fit for the low-mass T Tauri stars.
It is possible that MYSOs have A supergiant
configurations, from the results in Section \ref{sec:spt}. As the
stars are larger and hence the gravitational potential being released
per unit mass will be smaller, including this effect will thus result
in accretion rates that will be about an order of magnitude higher.

\begin{table}
\centering
\caption{Fraction of objects with different type of profile features. Objects can display more than one feature.}
\label{tab:profres}
\begin{tabular}{lllll}
\hline
                        & Br$\gamma$/Br12 &   \\
\hline \\
Flat                    & 19\%            &     \\
Wide peak               & 26\%            &   &  \\
Narrow peak             & 26\%            &   &  \\
Dish               & 19\%            &  &  \\
Dish + peak & 11\%            &   &  \\
Central dip             & 15\%            &  &  \\
Red asymmetry           & 15\%            &  &  \\
Blue asymmetry          & 15\%            &   & \\
\hline
\end{tabular}
\end{table}

\begin{figure}
\centering
\begin{tabular}{cc}
\includegraphics[scale=0.25]{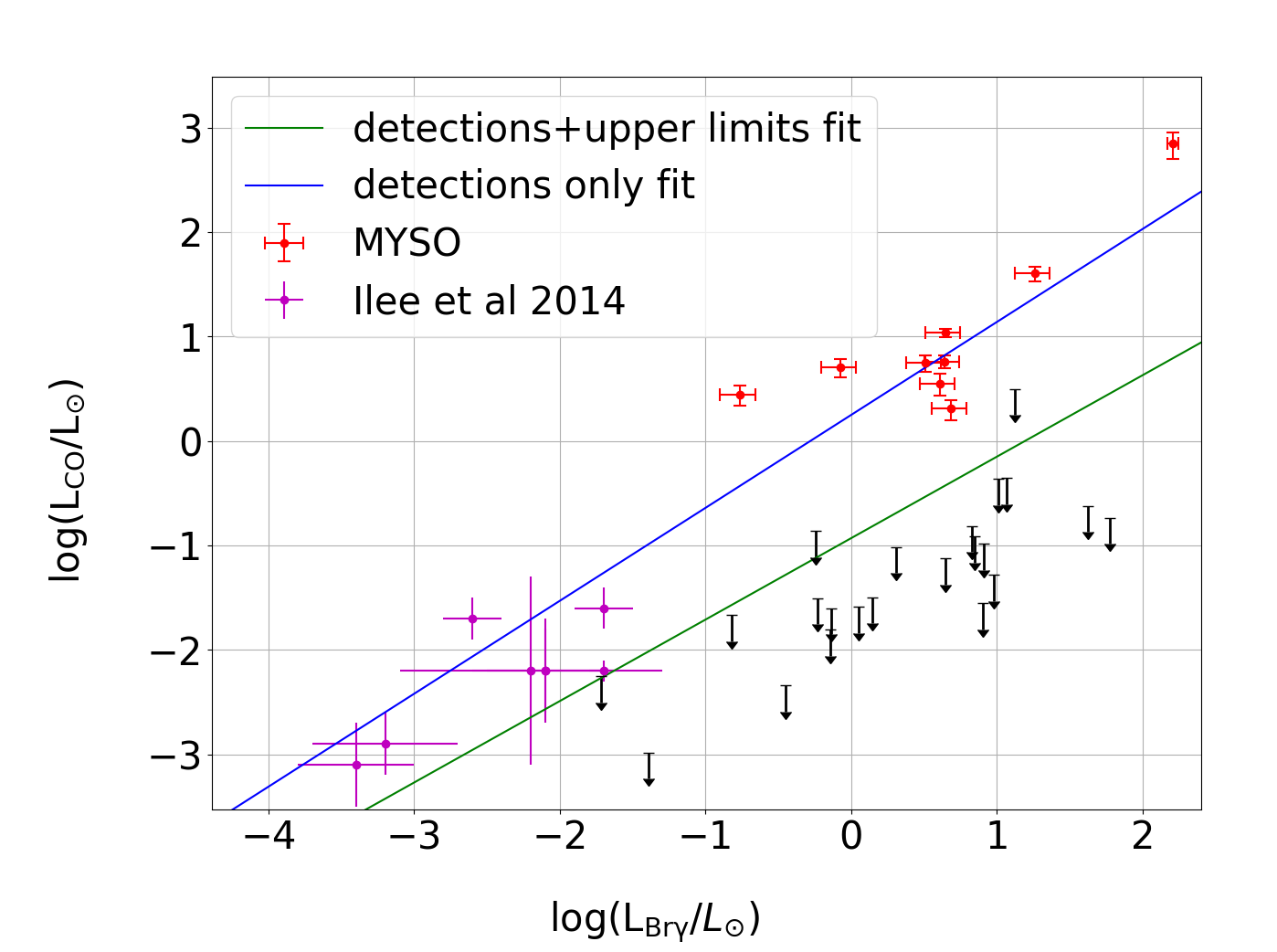}  
\end{tabular}
\caption{ Log-log plot of the CO bandhead luminosity as a function of Br$\gamma$ luminosity, with data from GNIRS and \citet{2014MNRAS.445.3723I}}
\label{fig:cobrg}
\end{figure}

\subsubsection{Other line luminosities}

Not only the Br12 and Br$\gamma$ line luminosities correlate (Figure
\ref{fig:br1210vg}), 75\% of the lines' luminosities studied here
correlate with one another with a probability of false correlation of
under 1\%.  As example, we find that CO bandhead, when in emission,
and Br$\gamma$ luminosities strongly correlate (see Figure
\ref{fig:cobrg}, in agreement with the results of
\citealt{2014MNRAS.445.3723I}). We find a best fit line to detections
of $\log(L_{CO})=(0.49\pm0.11)+(0.80\pm0.09)\log(L_{Br\gamma})$, and
when including upper limits, this becomes
$\log(L_{CO})=(-0.95\pm0.12)+(0.85\pm0.11)\log(L_{Br\gamma})$.

There are two issues that we need to highlight here. Firstly, the CO
bandhead emission arises from the warm, dense circumstellar disk
material which is not necessarily directly associated with the ionized
wind or present-day accretion flows. It is thus interesting that the luminosity of
emission lines originating from these very different regions are
correlated. A logical further step is that the correlation itself will
also allow one to derive the mass accretion rates from the CO bandhead
emission, despite the fact that the line forming region does not need
to have a direct relationship to the accretion process. Secondly, the
correlation is very strong when the CO bandhead is in emission, but a
large fraction of the stars in the sample does not have (detected)
emission. The upper limits to the line luminosities for the
non-detections lie in some cases several orders of magnitude below
those of the detections, which prompts the question why the detections
themselves would correlate with the Br$\gamma$ line luminosities at all.\\
A first conclusion of this would be that there may be an underlying
relation between these lines causing their luminosity to correlate. In
the case of the Br$\gamma$ line, which is more likely to be formed in
a wind rather than in an accretion funnel (see the discussion later on
line profiles), it could be argued that accretion and outflows are
intimately linked as winds can be accretion powered and thus
correlated with the accretion rates. However, many lines arise from
different regions and it will be hard to apply this explanation to all
lines. Indeed, in a recent study on Herbig Ae/Be and T Tauri stars,
\citet{2015MNRAS.452.2837M} have shown that the origin of line
correlations with accretion luminosities are caused by both of these
quantities being correlated with stellar luminosity.  This may well be
the case for MYSOs too, with correlations seen being a scaling effect
as the emission line strengths are proportional to the stellar
luminosity. Hence brighter, more massive, stars will have stronger emission
lines, as well as a stronger continuum flux.

\label{sec:lums}

\begin{figure*}
\centering
\includegraphics[scale=0.4]{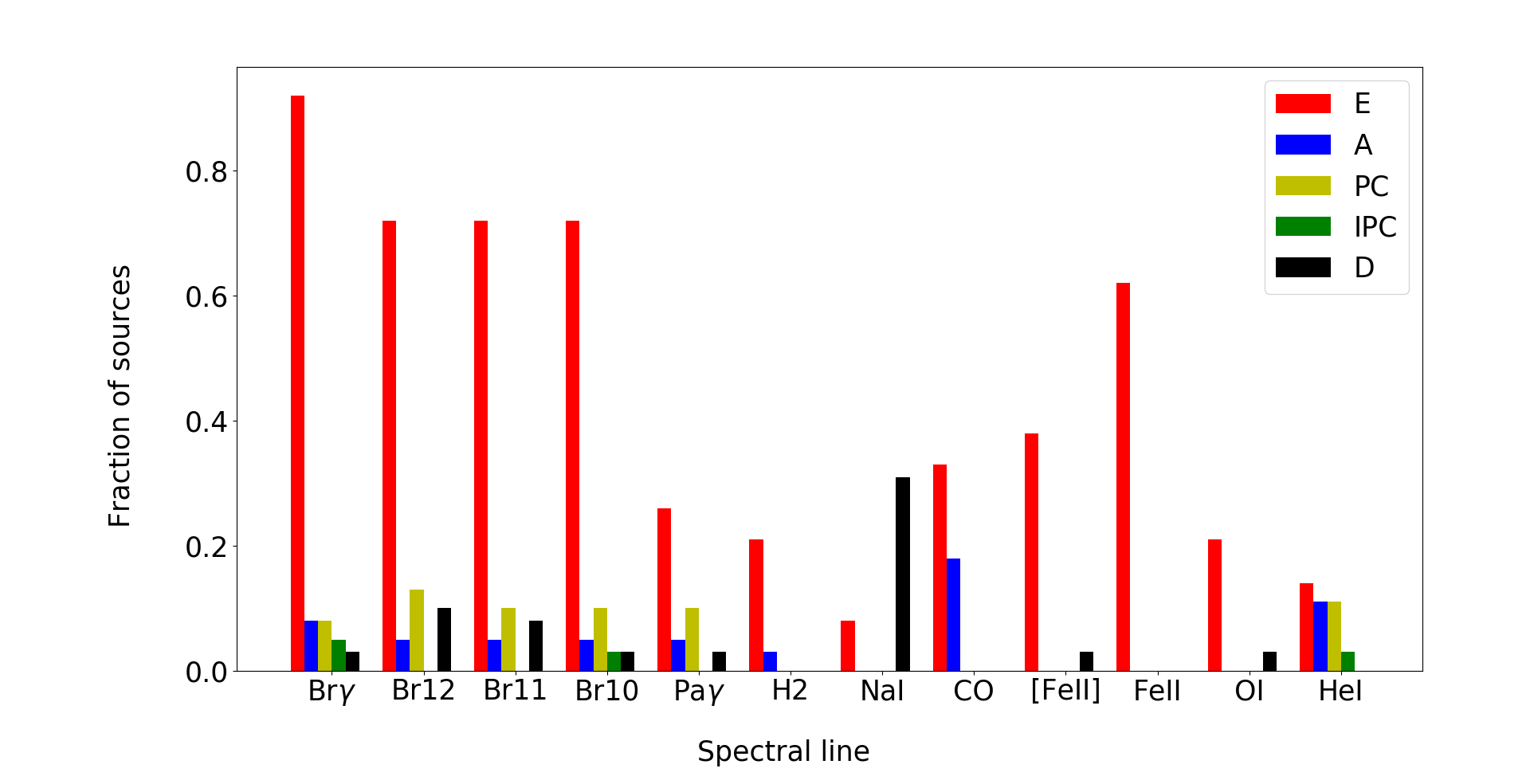}
\caption{ Detection rates of different features, with different profile
  types. E=single-peaked emission, A=absorption, PC=P Cygni profile,
  IPC=inverse P Cygni and D=double peaked}
\label{fig:atlas1}
\end{figure*}

\subsubsection{HII region contributions to recombination lines}

\begin{figure}
\centering
\begin{tabular}{cc}
\includegraphics[scale=0.25]{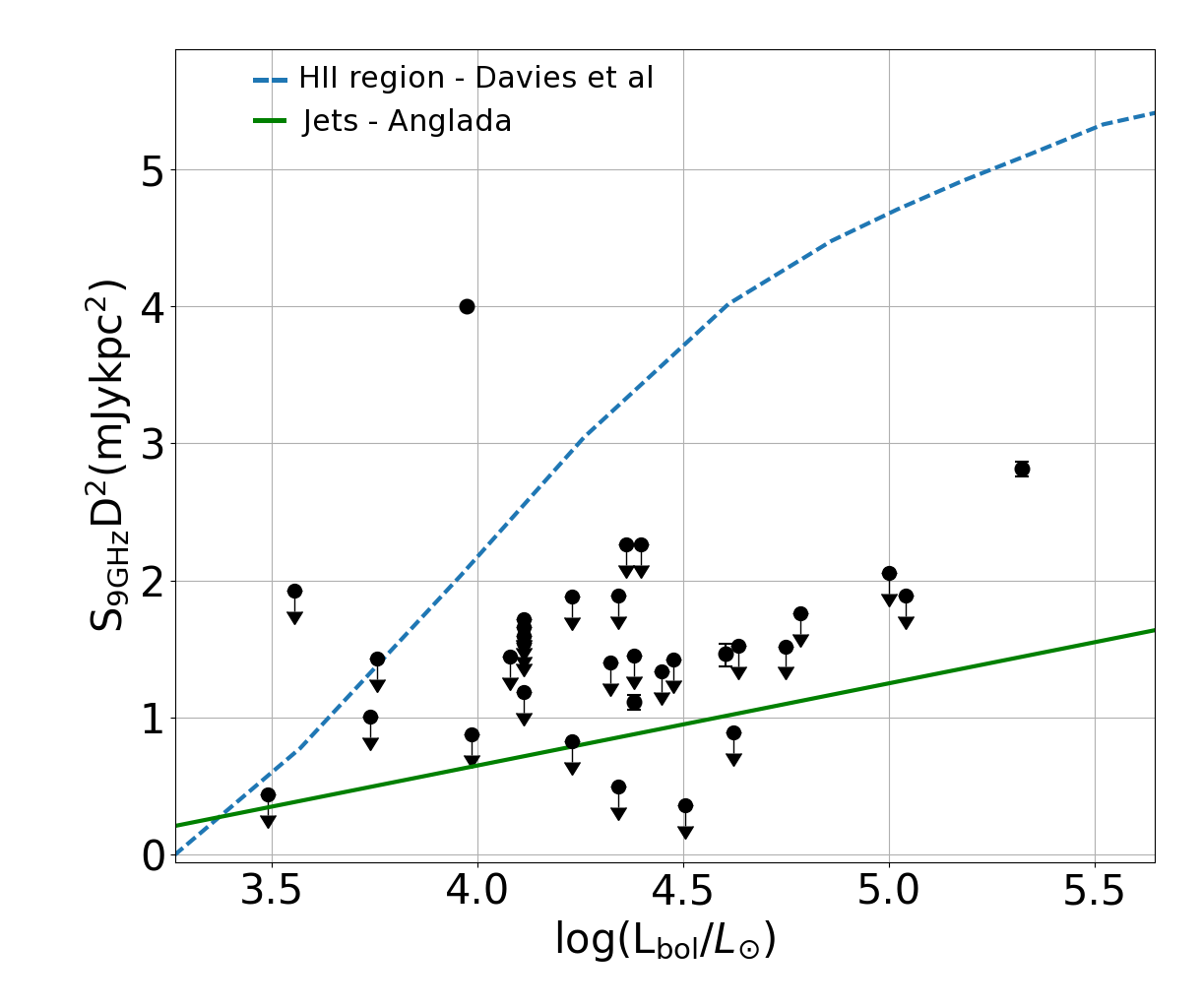}  
\end{tabular}
\caption{Radio flux at 9 GHz of our MYSOs, taken from the RMS database when available, as a function of bolometric luminosity. HII regions are expected to be found close to the dotted blue curve, as per the models of \citet{2011MNRAS.416..972D}, whereas MYSO jets closer to the green line, resulting from a fit to observations by \citet{anglada95}.}
\label{fig:hiijets}
\end{figure}
In the previous sections we attributed all of the Br$\gamma$ line luminosity to accretion. However, this assumes that there is no contribution from low density ionised gas like that of an HII region. Here we assess the plausibility of this assumption by analysing the possible presence of an HII region in the sample objects. \\
Firstly, one would expect a much stronger Br$\gamma$ line(4-5$\times$continuum level) than what we observe in our survey, where the average Br$\gamma$ strength is 1.3$\times$continuum. Recombination lines in HII regions are also narrower compared to our sample average of 150 km/s (typical FWHM of 30-40 km/s  e.g. \citealt{lumsden96}, \citealt{burton04}). \\
In addition, another means of revealing the presence of an HII region is by detecting strong free-free emission at 5 or 9 GHz. Most of our sources are undetected at these radio frequencies at the 5mJy level (\citealt{2009A&A...501..539U}, or in subsequent RMS database observations).\\
Several of our sources have been observed at high-resolution in the sensitive radio observations of \citet{rosero16}. Their morphologies show little evidence of an HII region structure. Instead, the radio continuum  is dominated by jet emission. Following \citet{purser16} we produce a plot of radio flux as a function of bolometric luminosity in Figure \ref{fig:hiijets} and we compare the MYSOs with expected values for HII regions from \citet{2011MNRAS.416..972D} and jets from \citet{anglada95}. We find that given their bolometric luminosity almost all of our sources are too faint in the radio to be an HII region. Four objects are detected at 9 GHz (G110.1082B, G076.3829, G111.5423 and G010.8411), with only G110.1082B being close to the HII region line. However, this MYSO is located 5" away from an HII region (G110.1082A), and the radio flux is integrated over both A and B sources. \\
Given the above, we conclude that the contribution of HII region emission to the total Br$\gamma$ flux is minimal.
\subsubsection{Profile shapes}
\begin{figure}
\centering
\begin{tabular}{cc}
\includegraphics[scale=0.3]{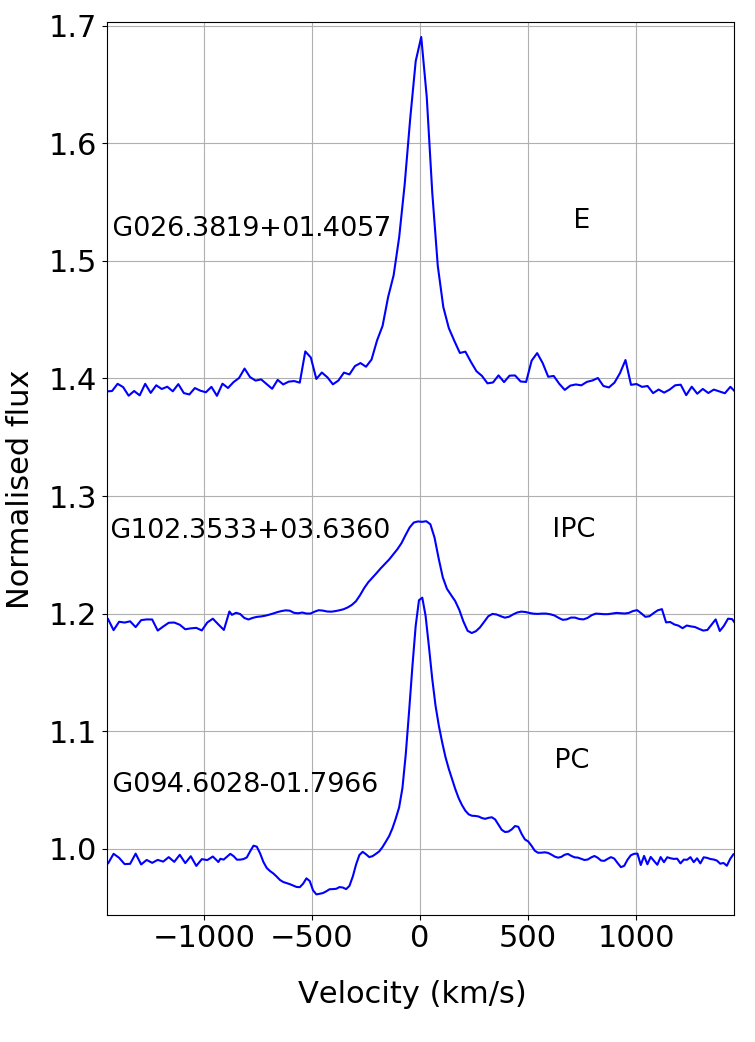}  
\end{tabular}
\caption{Examples of Br$\gamma$ profiles from our sample. From top to bottom: single-peaked emission G026.3819; inverse P Cygni absorption G102.3533; P Cygni absorption G094.6028}
\label{fig:brgprof}
\end{figure}

One of the main advantages of this data set is the spectral
resolution. This allows for a analysis of the line profiles.
Following the prescriptions of \citet{2015ApJ...810....5C}, who
performed an analysis of line profiles in Herbig Ae/Be stars, we
divide the profile features into 6 different categories: E -
single-peaked emission, A - absorption, PC - P Cygni profile, IPC -
inverse P Cygni profile, D - double peaked profile and N -
non-detection. The results are presented graphically, for the main
observed lines, in Figure \ref{fig:atlas1}.\\
For most lines, single-peaked emission is the most frequent feature
type.  The relative fractions of single peaked emission fractions to
other feature types are similar for all HI recombination
lines. Blueshifted P Cygni-type (PC) absorption, indicative of
outflowing material, is fairly rare; it is seen in all HI lines in
three MYSOs, in Br10-12 in two other objects, and in HeI in 4 objects.  Redshifted
inverse P Cygni-like (IPC) absorption, indicative of infall, is seen
in three objects in Br$\gamma$ and one in HeI. The occurrences of P
Cygni and inverse P Cygni lines are 13\% and 8\% respectively. These
are lower than the values found by \citet{2015ApJ...810....5C} for
Herbig Ae/Be stars, which in turn were found to be lower than those
for T Tauri stars. They also found that more Herbig Ae than Herbig Be
stars show IPC lines, which they take as indication of the differences
between the dominant accretion processes involved in these objects.\\
Our fractions of PC and IPC profiles show that this trend continues to
higher mass MYSOs. The PC velocities extend as far as 600-800 km/s,
and IPC up to 200-300 km/s.\\
An example of the Br$\gamma$ profiles can be seen in Figure \ref{fig:brgprof}.

\label{sec:profsh}

\subsubsection{Hydrogen line profile ratios}
\label{sec:profiles}

As shown in Section \ref{sec:profsh}, many observed HI recombination
line profiles show asymmetries, suggesting that there are multiple
components giving rise to the observed emission. In order to
disentangle these, we analyse the recombination line ratios. These can
provide clues on the formation mechanism and kinematics of wind
emission. As explained earlier, the value of the
Br$\gamma$/Br12 line ratio increases with decreasing optical
depth. The spectral dependence of this ratio can help us distinguish
between different wind components of different accelerations giving
rise to the observed emission. In the earlier work of
\citet{1993MNRAS.265...12D} and \citet{1995MNRAS.272..346B} ratios of
Br$\alpha$/Br$\gamma$ and Br$\alpha$/Pf$\gamma$ were studied.
Our Br$\gamma$/Br12 ratios are expected to be comparable with those
as the absorption cross sections of these lines have similar values to
those of the lines they used (see also \citealt{2012MNRAS.424.1088L}).
\begin{figure*}
\centering
\includegraphics[scale=0.18]{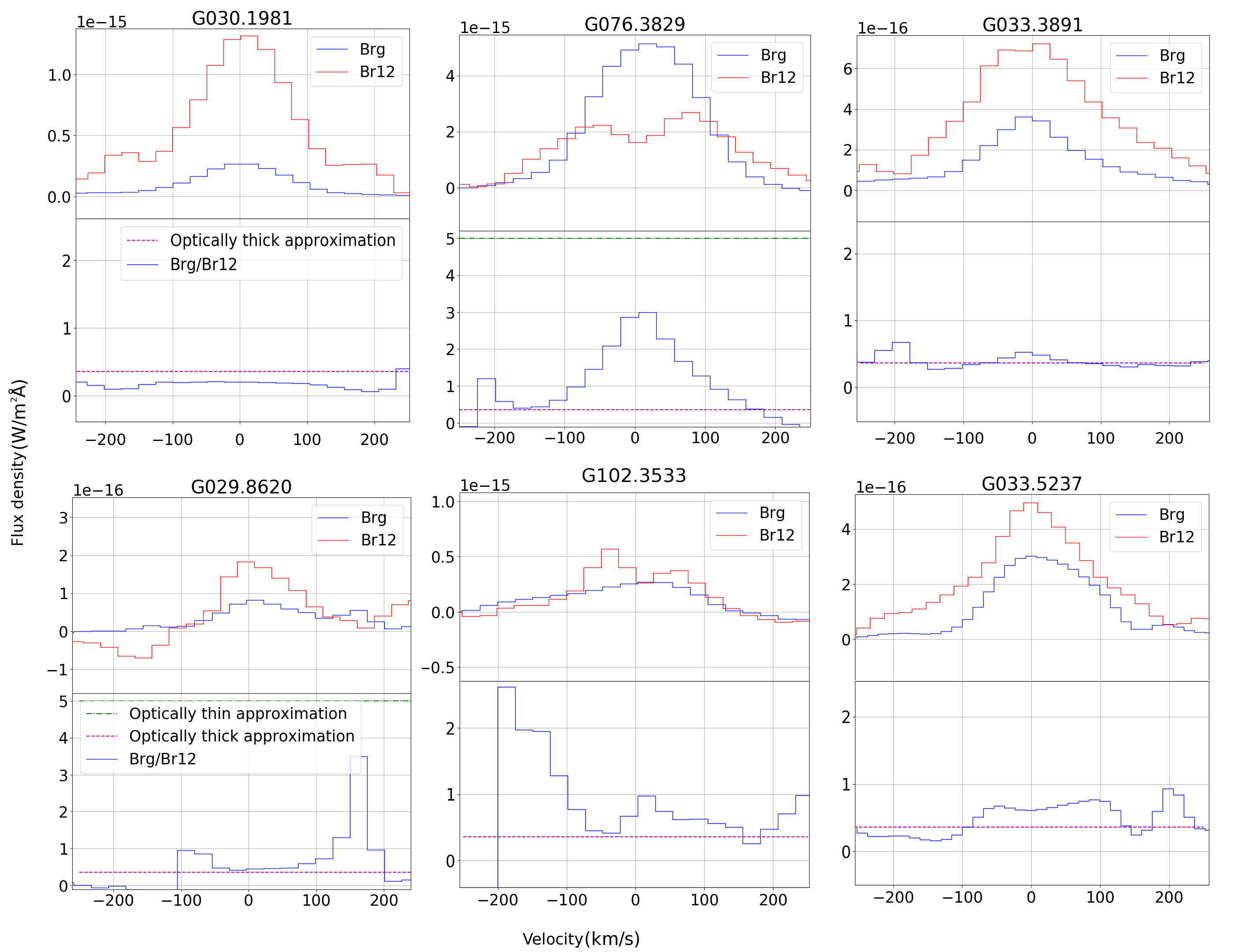} 

\caption{Examples of different types of profile ratio in
  Br$\gamma$/Br12. The horizontal axis denotes the velocity. Top (left to right): Flat ratio, Wide peak, Narrow
  peak. Bottom(left to right): Dish shape, Dish shape+central
  peak, Central dip.}

\label{fig:profratios}
\end{figure*}

The procedure for computing the ratios is as follows. We select a
region of $\pm$500 km/s from the velocity center of the line, and
rebin the spectra to correct for differences in resolution between the
$K$ and $H$ bands. The extinction correction is applied and the continuum is fitted with a 2$^{\rm nd}$ order polynomial
fit and subtracted, and the two lines are divided. The chosen
extinction value only affects the numerical value of the ratio, but
that trends in shape of the ratio profile remain unchanged.  A number
of different features can be distinguished in the ratios, as shown in
Figure \ref{fig:profratios}. Some line ratio profiles show flat wings
and a core, others a dish shaped profile. Central peaks are also seen,
either narrow (FWHM$\approx$50-80 km/s) or wide (FWHM$\approx$80-100
km/s). Other objects display a central dip, blue or red
asymmetries. Often two or more of these different types are all seen
in a line ratio. Their relative proportions are shown in Table
\ref{tab:profres}.\\
Some of these features have been seen in other line ratio
surveys. \citet{1993MNRAS.265...12D} reported a dish shaped profile in
S106IRS1 in their Br$\alpha$/Br$\gamma$ spectral line ratio, which
they interpret as an accelerated wind, while
\citet{2012MNRAS.424.1088L} find a peaked shape in most of the
Br$\gamma$/Br12 ratios.\\
A dish shaped profile will indicate optically thin gas expanding
faster than the optically thick central
material. \citet{1993MNRAS.265...12D} also see a narrow central peak
in the S106IRS1 ratio, which they interpret as an optically thin
nebular gas region. They find single peaks in both Br$\alpha$ and Br$\gamma$. We also see a single peak in Br$\gamma$, but double peaked Br12. In addition to the wind, this could indicate the presence of a rotating disc. This is in agreement with radio observations of \citet{gibb07}, who find an elongated structure, which they identify with an equatorial disk wind. However, the disk is probably not fully ionised given the detection of CO bandhead emission towards this object.\\ 
By the same logic, a weak and wide peak can be
seen as an inverted dish shape, where the optically thin gas is
expanding slower than the optically thick gas. We interpret this as a
decelerating wind. A central dip within a wider peak can be seen as a
decelerating wind combined with a larger, optically thick nebular
region.  A flat ratio simply shows all regions expanding at the same
acceleration, irrespective of optical thickness. Asymmetries are
suggestive of infall (on the red side) or outflow (on the blue side).\\
Regarding any correlations between the HI ratio features and
bolometric luminosity, we find that there is no clear trend between
these two quantities. We also found no correlation with the \citet{coopthesis} MYSO
Types 1-3, or with 2MASS NIR colours. This may be due to our small
sample size. It is possible that with a distance-limited full
sample such trends might appear. Still, it is interesting to note the
variety of features observed. Following our presence of absorption
lines in a number of our targets (see Section \ref{sec:spt}), we also
investigated the effect of the photospheric absorption lines on the
observed Br$\gamma$/Br12 wind line ratios. The emission first fills in
the photospheric absorption lines and arguably this needs to be taken
into account in case the continuum excess emission is small. We worked
this out for a range of extinction and excess values and find that the effect induced by photospheric absorption lines will be less than 10\% of the Br$\gamma$/Br12
ratio value in the wings.  We leave a detailed analysis of this effect
for all our targets to a future study, but it is worth remembering
that core features in the wind line ratios that are less than 10\% of
the wing value are likely caused by intrinsic photospheric absorption.

\section{Conclusions}
\label{sec:concl}

We have presented medium-resolution echelle NIR spectroscopic data for
36 MYSOs from the GNIRS instrument on the Gemini-North
telescope \footnote{
\textbf{Note} - The spectral atlas with equivalent widths, line fluxes and luminosities as well as continuum-normalised full spectra are available as appendices which can be found only in the online version of this paper.}. Nearly a third of these constitute the first NIR spectra
taken for these objects. This is the largest sample of MYSOs studied at this
resolution at these wavelengths of MYSOs to date. Dust extinction is
estimated from the slope of the continuum in the $H$ band. Our main
findings are as follows:

\begin{itemize}

\item We have found photospheric absorption lines for one object,
  G015.1288. These lines indicate that this source is best fit by a
  spectral type of AIb. An MYSO with an A supergiant configuration is
  consistent with the simulations of swollen up MYSOs of
  \citet{2010ApJ...721..478H} and \citet{2016MNRAS.458.3299H}.

\item The accretion luminosities and rates from Br$\gamma$ agree with
  results for lower mass sources, providing tentative evidence for a
  continuity of star formation processes across a large mass range.

\item 75\% of the lines of interest studied in this survey correlate
  with one another in terms of luminosity. This may be due to
  correlations with stellar luminosity.

\item Br$\gamma$/Br12 line profile ratios show a wide variety of
  features, possibly corresponding to a range of winds. We see ratios
  corresponding to accelerated, constant or decelerated wind
  characteristics, as well as traces of wider regions of different
  optical thickness.

\item We find no correlation between line ratio features and
  luminosity, NIR colours or MYSO Types cf. \citet{coopthesis}

\end{itemize}

\section*{Acknowledgements}

RP gratefully acknowledges the studentship funded by the Science and Technologies Facilities Council of the United Kingdom. We thank the anonymous referee for the comments and suggestions, which improved the quality of this article. This publication makes use of data products from the Two Micron All
Sky Survey, which is a joint project of the University of
Massachusetts and the Infrared Processing and Analysis
Center/California Institute of Technology, funded by the National
Aeronautics and Space Administration and the National Science
Foundation. We also make use of the SIMBAD data base, operated at CDS, Strasbourg,
France. This paper made use of information from the Red MSX Source
survey database at
\url{http://rms.leeds.ac.uk/cgi-bin/public/RMS_DATABASE.cgi} which was
constructed with support from the Science and Technology Facilities
Council of the UK. The reduction process made use of PyRAF, which is a
product of the Space Telescope Science Institute, operated by AURA for
NASA.

\bibliographystyle{mn2e} 
\bibliography{bibliography_feb2017}

\appendix 

\begin{landscape}
\begin{table}
\centering
\caption{Spectral atlas, with equivalent widths in {\AA}, fluxes uncorrected for extinction and in W/m$^{2}${\AA} and luminosities in L$_{\odot}$}
\label{atlas_table1}
\begin{tabular}{llllllllllllllll}
\hline
Object                 & Upper limit EW & EW$_{Br\gamma}$ & $\Delta$EW$_{Br\gamma}$ & F$_{Br\gamma}$ & dF$_{Br\gamma}$ & L$_{Br\gamma}$ & $\Delta$ L$_{Br\gamma}$ & L$_{Acc}$ & $\Delta$ L$_{Acc}$ & M    & $\Delta$ M & R    & $\Delta$ R & $\dot{M}$ & $\Delta \dot{M}$ \\
\hline
G010.8411-02.5919      & 0.09           & -0.6            & 0.2                     & 5.4E-18        & 1.4E-18         & 1.5E-01        & 4.1E-02                 & 6.5E+02   & 1.8E+02            & 15.9 & 0.8      & 6.0  & 0.3      & 7.9E-06    & 3.9E-07     \\
G010.8856+00.1221      & 0.10           & -1.8            & 0.2                     & 1.1E-17        & 1.0E-18         & 5.9E-01        & 1.6E-01                 & 2.2E+03   & 5.9E+02            & 10.5 & 0.5      & 2.9  & 0.1      & 1.9E-05   & 9.7E-07     \\
G012.9090-00.2607      & 0.11           & -2.3            & 0.2                     & 2.1E-17        & 1.7E-18         & 1.0E-01        & 2.7E-02                 & 4.5E+02   & 1.2E+02            & 17.2 & 0.9      & 6.9  & 0.3      & 4.5E-03    & 2.2E-04     \\
G014.9958-00.6732      & 0.07           & -18.1           & 0.1                     & 9.6E-16        & 6.2E-18         & 8.0E+00        & 2.1E+00                 & 2.3E+04   & 6.2E+03            & 13.4 & 0.7      & 4.4  & 0.2      & 2.4E-04    & 1.2E-05    \\
G015.1288-00.6717      & 0.05           & -1.9            & 0.1                     & 2.3E-17        & 1.0E-18         & 4.1E-02        & 1.1E-02                 & 2.0E+02   & 5.3E+01            & 13.1 & 0.7      & 4.2  & 0.2      & 2.1E-06    & 1.0E-07     \\
G017.6380+00.1566      & 0.01           & -0.6            & 0.0                     & 2.9E-17        & 1.3E-18         & 4.4E-02        & 1.2E-02                 & 1.5E+02   & 4.1E+01            & 8.5  & 0.2      & 3.2  & 0.1      & 3.1E-05   & 8.4E-07     \\
\hline
\end{tabular}
\end{table}  

\clearpage
\newpage

\begin{table}
\centering
\caption{Spectral atlas, with equivalent widths in {\AA}, fluxes uncorrected for extinction and in W/m$^{2}${\AA} and luminosities in L$_{\odot}$}
\label{atlas_table2}
\begin{tabular}{lllllllllllll}
\hline
Object                 & EW$_{CO}$              & $\Delta$EW$_{CO}$ & F$_{CO}$ & $\Delta$ F$_{CO}$ & L$_{CO}$ & $\Delta$ L$_{CO}$ & EW$_{Br12}$ & $\Delta$EW$_{Br12}$ & F$_{Br12}$ & $\Delta$ F$_{Br12}$ & L$_{Br12}$ & $\Delta$ L$_{Br12}$ \\    
\hline  
G010.8411-02.5919      &                   &          &                 &          &                   &             &                     &            &                   &            &                     &         \\
G010.8856+00.1221      &                   &          &                 &          &                   &             & -0.5                & 0.1        & 3.5E-19           & 3.3E-20    & 3.5E-01             & 1.3E-01 \\
G012.9090-00.2607      & -5.5              & 0.3      & 1.2E-16         & 5.2E-18  & 7.0E+02           & 6.0E+01     &                     &            &                   &            &                     &         \\
G014.9958-00.6732      &                   &          &                 &          &                   &             & -12.5               & 0.1        & 1.6E-16           & 1.1E-18    & 1.1E+01             & 4.3E+00 \\
G015.1288-00.6717      &                   &          &                 &          &                   &             & 3.0                 & 0.1        & -2.4E-17          & 8.1E-19    & 1.5E-01             & 5.8E-02 \\
G017.6380+00.1566      &                   &          &                 &          &                   &             &                     &            &                   &            &                     &         \\
G018.3412+00.1566      &                   &          &                 &          &                   &             & -1.7                & 0.1        & 1.1E-18           & 4.5E-20    & 5.3E+00             & 2.1E+00 \\
\hline
\end{tabular}
\end{table}

\clearpage
\newpage

\begin{table}
\centering
\caption{Spectral atlas, with equivalent widths in {\AA}, fluxes uncorrected for extinction and in W/m$^{2}${\AA} and luminosities in L$_{\odot}$}
\label{atlas_table3}
\begin{tabular}{lllllllllllll}
\hline
Object                 & EW$_{Br11}$($\AA$) & $\Delta$EW$_{Br11}$ & F$_{Br11}$ & $\Delta$ F$_{Br11}$ & L$_{Br11}$ & $\Delta$ L$_{Br11}$ & EW$_{Br10}$ & $\Delta$EW$_{Br10}$ & F$_{Br10}$ & $\Delta$ F$_{Br10}$ & L$_{Br10}$ & $\Delta$ L$_{Br10}$ \\
\hline
G010.8411-02.5919      &                    &                     &            &                   &            &                     &             &                     &            &                   &            &                     \\
G010.8856+00.1221      & -0.5               & 0.2                 & 3.8E-19    & 1.5E-19           & 4.5E-01    & 1.8E-01             & -0.1        & 0.1                 & 1.5E-19    & 6.0E-20           & 6.6E-01    & 2.6E-01             \\
G012.9090-00.2607      &                    &                     &            &                   &            &                     &             &                     &            &                   &            &                     \\
G014.9958-00.6732      & -13.7              & 5.3                 & 1.8E-16    & 7.1E-17           & 1.3E+01    & 5.2E+00             & -17.2       & 6.7                 & 2.8E-16    & 1.1E-16           & 1.4E+01    & 5.6E+00             \\
G015.1288-00.6717      & 2.7                & 1.0                 & 1.8E-17    & 7.0E-18           & 1.5E-01    & 6.0E-02             & 3.4         & 1.3                 & 2.4E-17    & 9.4E-18           & 2.4E-01    & 9.5E-02             \\
G017.6380+00.1566      &                    &                     &            &                   &            &                     &             &                     &            &                   &            &                     \\
G018.3412+00.1566      & -2.0               & 0.8                 & 1.5E-18    & 6.0E-19           & 7.9E+00    & 3.1E+00             & -1.7        & 0.7                 & 1.9E-18    & 7.3E-19           & 1.4E+01    & 5.5E+00             \\
\hline         
\end{tabular}
\end{table}

\clearpage
\newpage

\begin{table}
\centering
\caption{Spectral atlas, with equivalent widths in Angstroms, fluxes uncorrected for extinction and in W/m$^{2}$ {\AA} and luminosities in solar luminosities}
\label{atlas_table4}
\begin{tabular}{lllllllllllll}
\hline
Object                 & EW$_{NaI}$ & $\Delta$EW$_{NaI}$ & F$_{NaI}$ & $\Delta$ F$_{NaI}$ & L$_{NaI}$ & $\Delta$ L$_{NaI}$ & EW$_{H_{2}}$ & $\Delta$EW$_{H_{2}}$ & F$_{H_{2}}$ & $\Delta$ F$_{H_{2}}$ & L$_{H_{2}}$ & $\Delta$ L$_{H_{2}}$ \\
\hline
G010.8411-02.5919      &            &                    &           &                  &           &                    &              &                      &             &                   &             &                      \\
G010.8856+00.1221      &            &                    &           &                  &           &                    &              &                      &             &                   &             &                      \\
G012.9090-00.2607      & -0.5      & 0.2               & 7.7E-18   & 2.1E-18          & 5.1E+01   & 1.4E+01            &              &                      &             &                   &             &                      \\
G014.9958-00.6732      &            &                    &           &                  &           &                    & -0.2         & 0.1                  & 1.4E-17     & 5.5E-18           & 1.8E-03     & 6.9E-04              \\
G015.1288-00.6717      &            &                    &           &                  &           &                    &              &                      &             &                   &             &                      \\
G017.6380+00.1566      &            &                    &           &                  &           &                    &              &                      &             &                   &             &                      \\
G018.3412+00.1566      &            &                    &           &                  &           &                    &              &                      &             &                   &             &                      \\
\hline
\end{tabular}
\end{table}

\clearpage
\newpage

\begin{table}
\caption{Spectral atlas, with equivalent widths in {\AA}, fluxes uncorrected for extinction and in W/m$^{2}${\AA} and luminosities in L$_{\odot}$}
\label{atlas_table6}
\begin{tabular}{lllllllllllll}
\hline
Object                 & EW$_{HeI}$ & $\Delta$EW$_{HeI}$ & F$_{HeI}$ & $\Delta$ F$_{HeI}$ & L$_{HeI}$ & $\Delta$ L$_{HeI}$ & EW$_{OI}$ & $\Delta$EW$_{OI}$ & F$_{OI}$ & $\Delta$ F$_{OI}$ & L$_{OI}$ & $\Delta$ L$_{OI}$ \\
\hline
G010.8411-02.5919      &            &                    &           &                  &           &                    &           &                   &          &                 &          &                   \\
G010.8856+00.1221      &            &                    &           &                  &           &                    &           &                   &          &                 &          &                   \\
G012.9090-00.2607      &            &                    &           &                  &           &                    &           &                   &          &                 &          &                   \\
G014.9958-00.6732      &            &                    &           &                  &           &                    & -15.5     & 4.2               & 1.1E-16  & 5.6E-18         & 2.6E+02  & 1.3E+01           \\
G015.1288-00.6717      & 7.1        & 0.3                & 4.3E-18   & 1.9E-19          & 5.3E-04   & 1.4E-04            &           &                   &          &                 &          &                   \\
G017.6380+00.1566      &            &                    &           &                  &           &                    &           &                   &          &                 &          &                   \\
G018.3412+00.1566      &            &                    &           &                  &           &                    &           &                   &          &                 &          &                   \\
\hline
\end{tabular}
\end{table}

\clearpage
\newpage

\begin{table}
\centering
\caption{Spectral atlas, with equivalent widths in {\AA}, fluxes uncorrected for extinction and in W/m$^{2}${\AA} and luminosities in L$_{\odot}$}
\label{atlas_table7}
\begin{tabular}{lllllll}
\hline
Object                 & EW$_{Pag}$ & $\Delta$EW$_{Pag}$ & F$_{Pa \gamma}$ & $\Delta$ F$_{Pa \gamma}$ & L$_{Pa \gamma}$ & $\Delta$ L$_{Pa \gamma}$ \\
\hline
G010.8411-02.5919      &            &                    &                 &                        &                 &                          \\
G010.8856+00.1221      &            &                    &                 &                        &                 &                          \\
G012.9090-00.2607      &            &                    &                 &                        &                 &                          \\
G014.9958-00.6732      &            &                    &                 &                        &                 &                          \\
G015.1288-00.6717      & -0.8       & 0.3                & 4.7E-19         & 1.8E-19                & 1.0E+00         & 3.9E-01                  \\
G017.6380+00.1566      &            &                    &                 &                        &                 &                          \\
G018.3412+00.1566      &            &                    &                 &                        &                 &                          \\
\hline
\end{tabular}
\end{table}

\clearpage
\newpage

\begin{figure}
\includegraphics[scale=0.6]{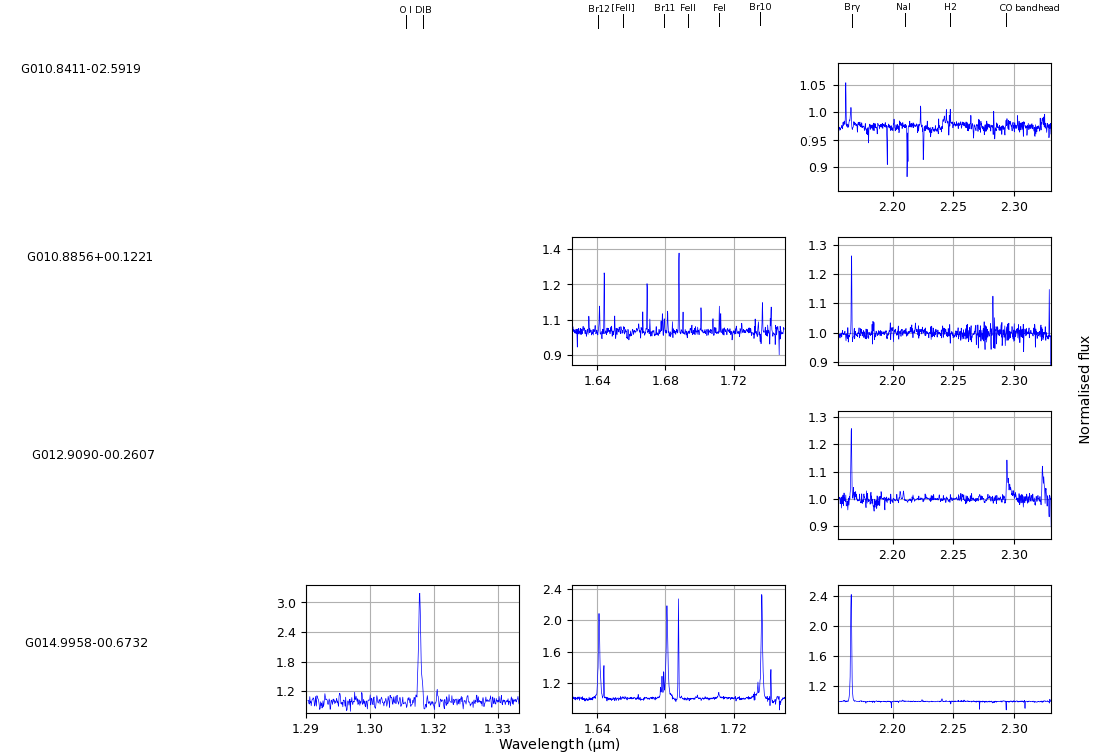}
\caption{Continuum normalised spectra, with left to right panels showing X, J, H and K bands}
\label{fig:sp1}
\end{figure}

\clearpage 
 
\begin{figure}
\includegraphics[scale=0.6]{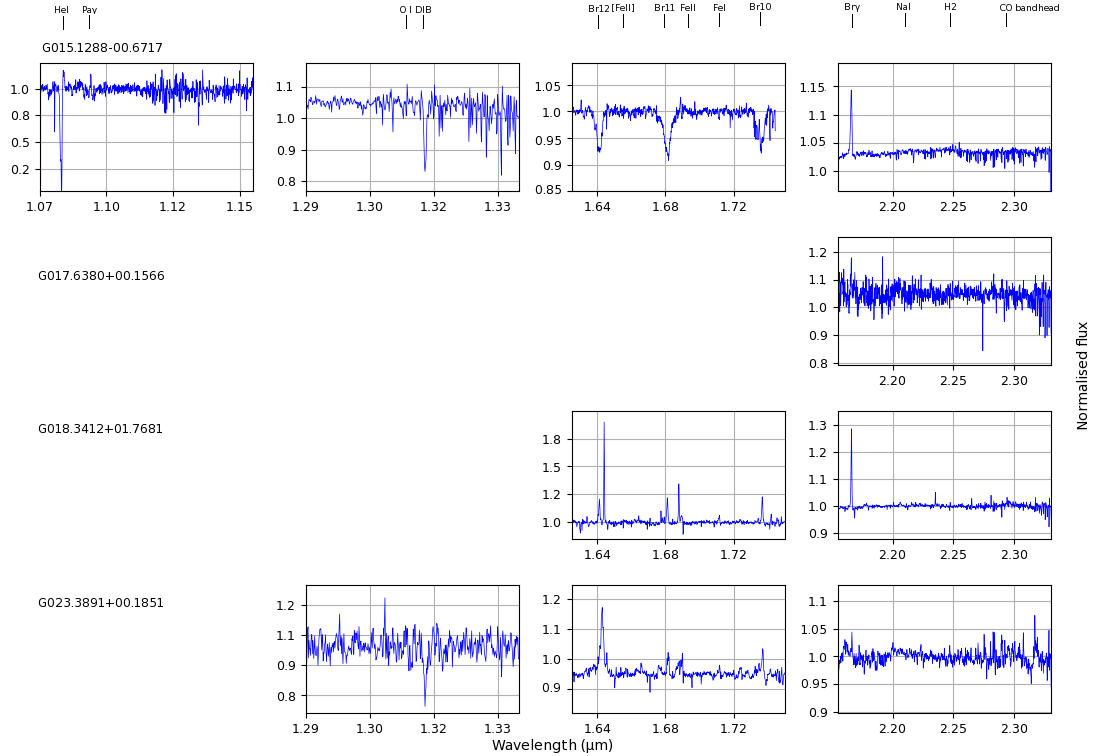}
\caption{Continuum normalised spectra, with left to right panels showing X, J, H and K bands}
\label{fig:sp2}
\end{figure}

\clearpage 
 
\begin{figure}
\includegraphics[scale=0.6]{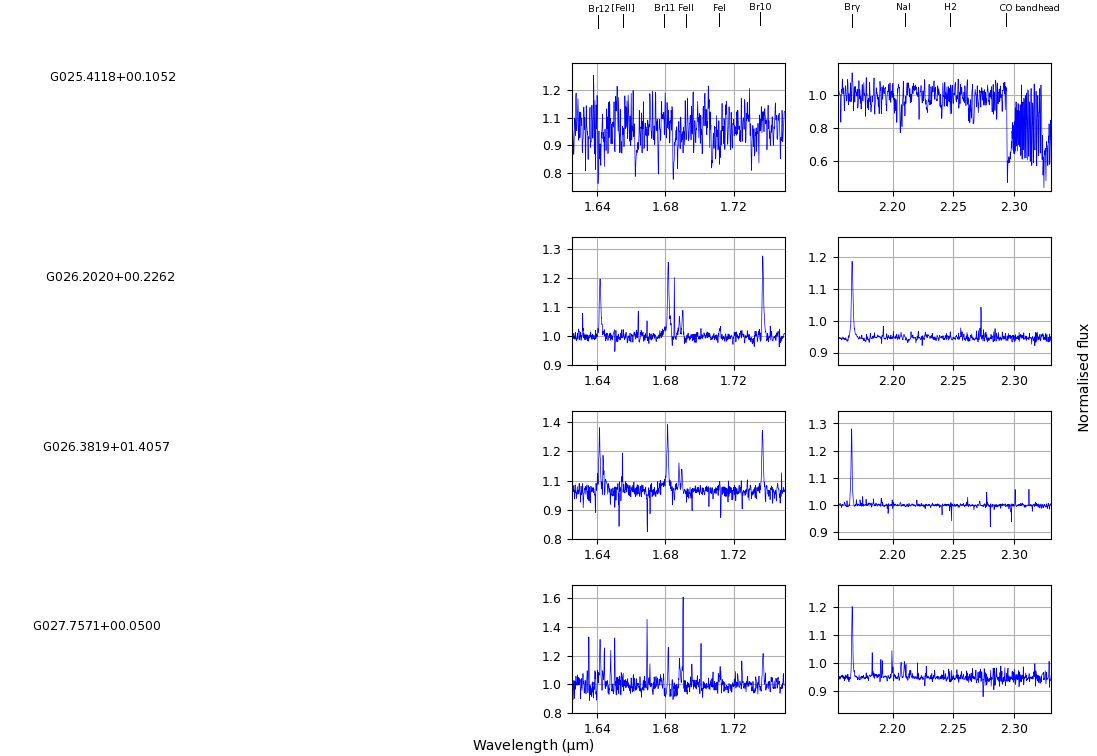}
\caption{Continuum normalised spectra, with left to right panels showing X, J, H and K bands}
\label{fig:sp3}
\end{figure}

\clearpage 
 
\begin{figure}
\includegraphics[scale=0.6]{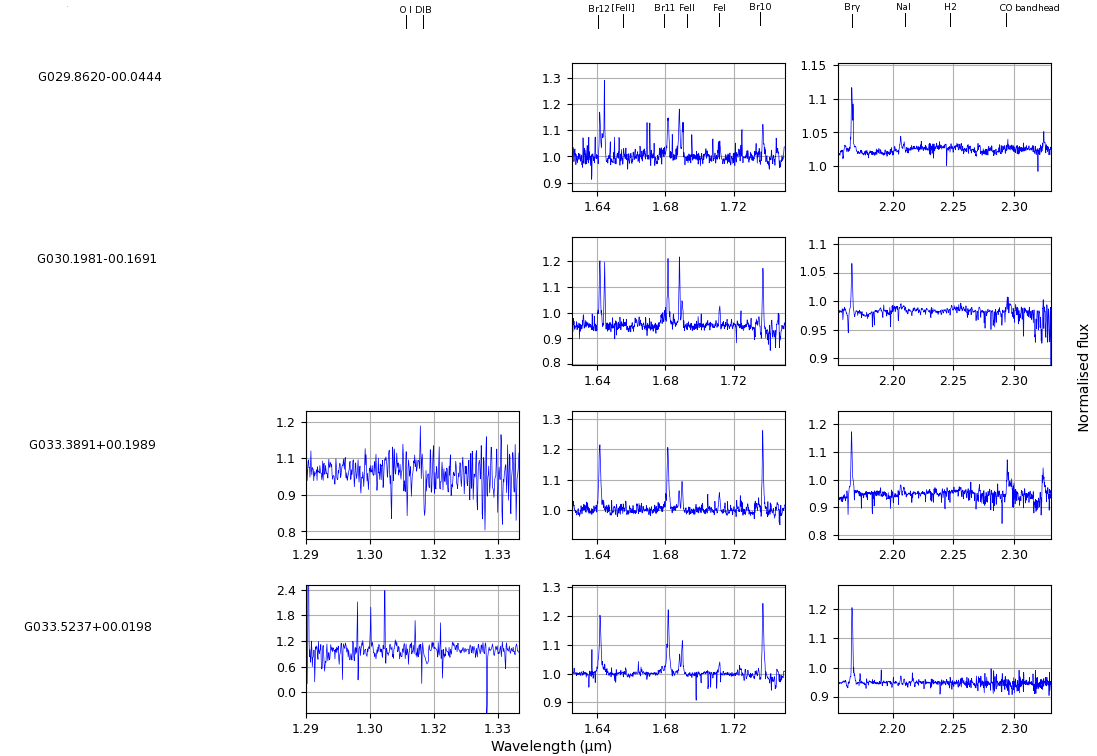}
\caption{Continuum normalised spectra, with left to right panels showing X, J, H and K bands}
\label{fig:sp4}
\end{figure}
\clearpage 
 
\begin{figure}
\includegraphics[scale=0.6]{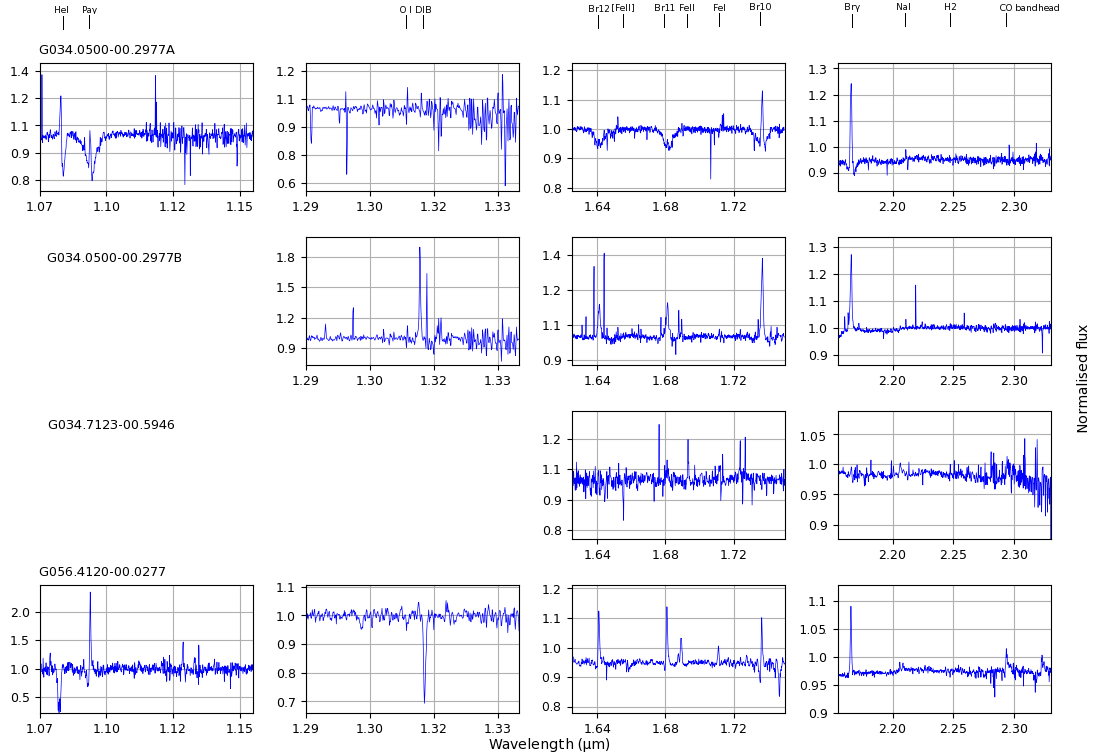}
\caption{Continuum normalised spectra, with left to right panels showing X, J, H and K bands}
\label{fig:sp5}
\end{figure}

\clearpage 
  
\begin{figure}
\includegraphics[scale=0.6]{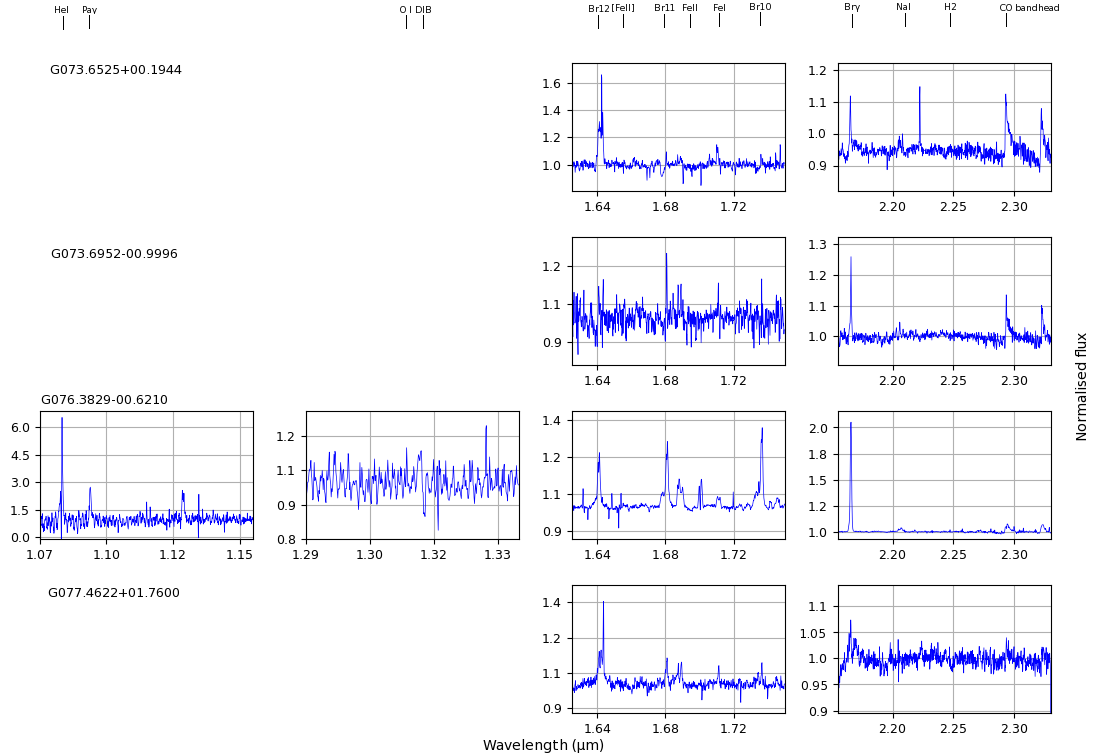}
\caption{Continuum normalised spectra, with left to right panels showing X,J,H and K bands}
\label{fig:sp6}
\end{figure}

\clearpage 
 
\begin{figure}
\includegraphics[scale=0.6]{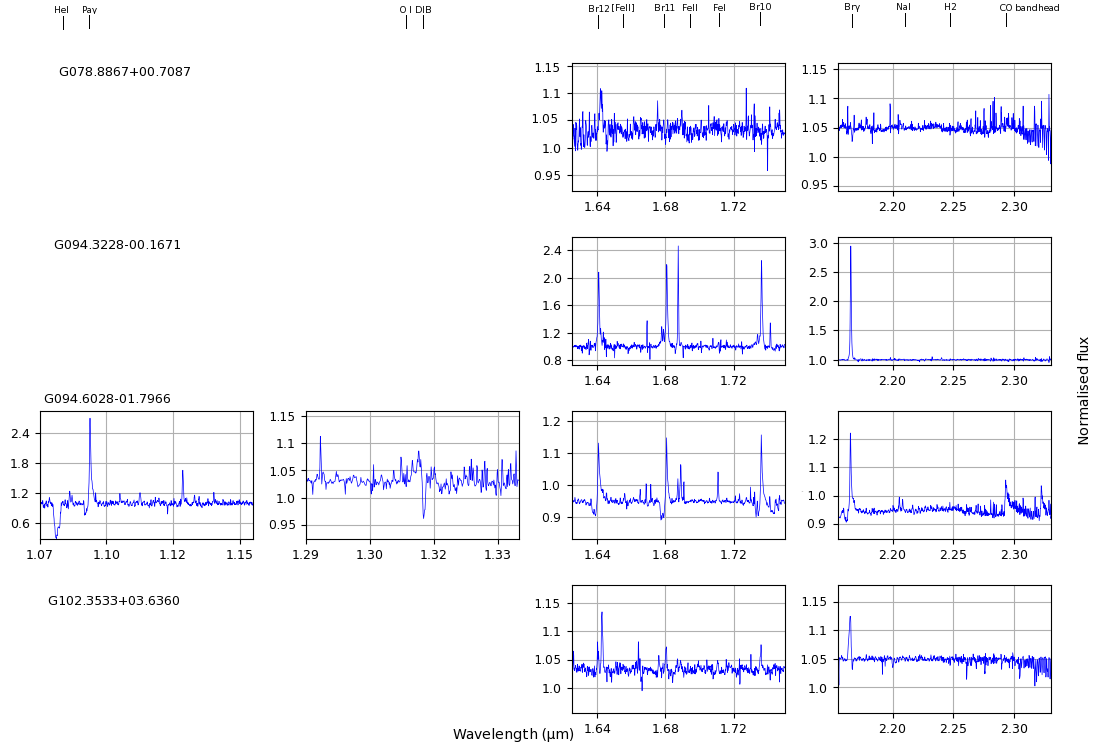}
\caption{Continuum normalised spectra, with left to right panels showing X, J, H and K bands}
\label{fig:sp7}
\end{figure}

\clearpage 
 
\begin{figure}
\includegraphics[scale=0.6]{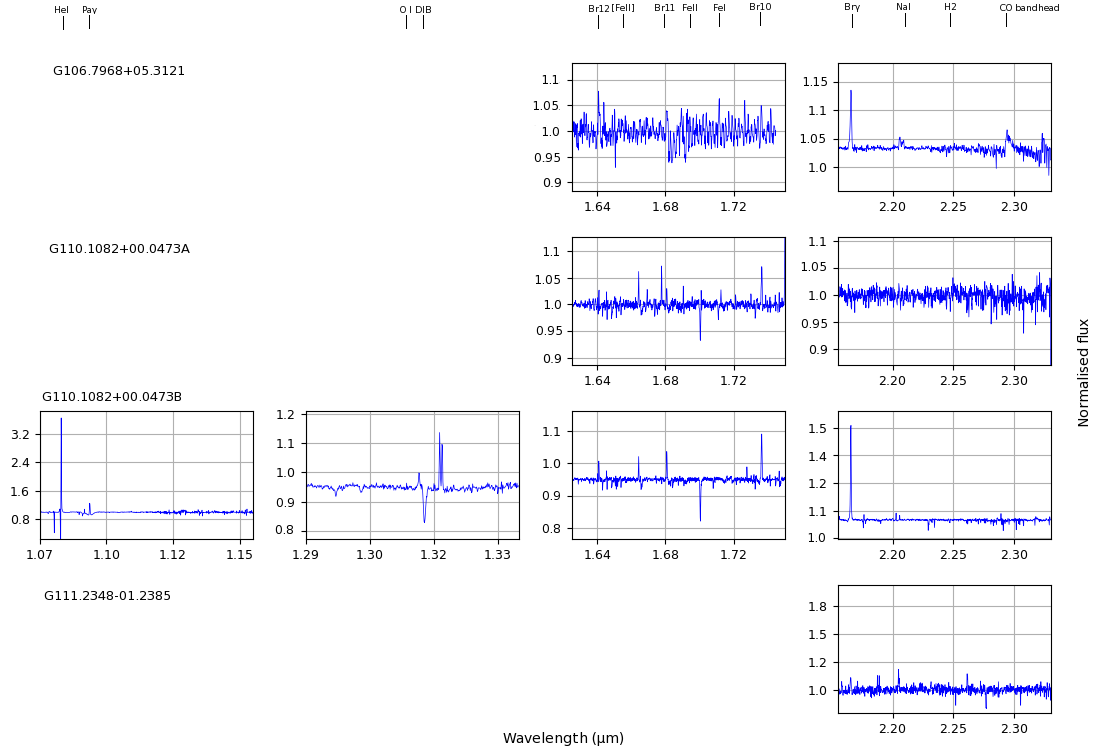}
\caption{Continuum normalised spectra, with left to right panels showing X,J,H and K bands}
\label{fig:sp8}
\end{figure}

\clearpage 
 
\begin{figure}
\includegraphics[scale=0.6]{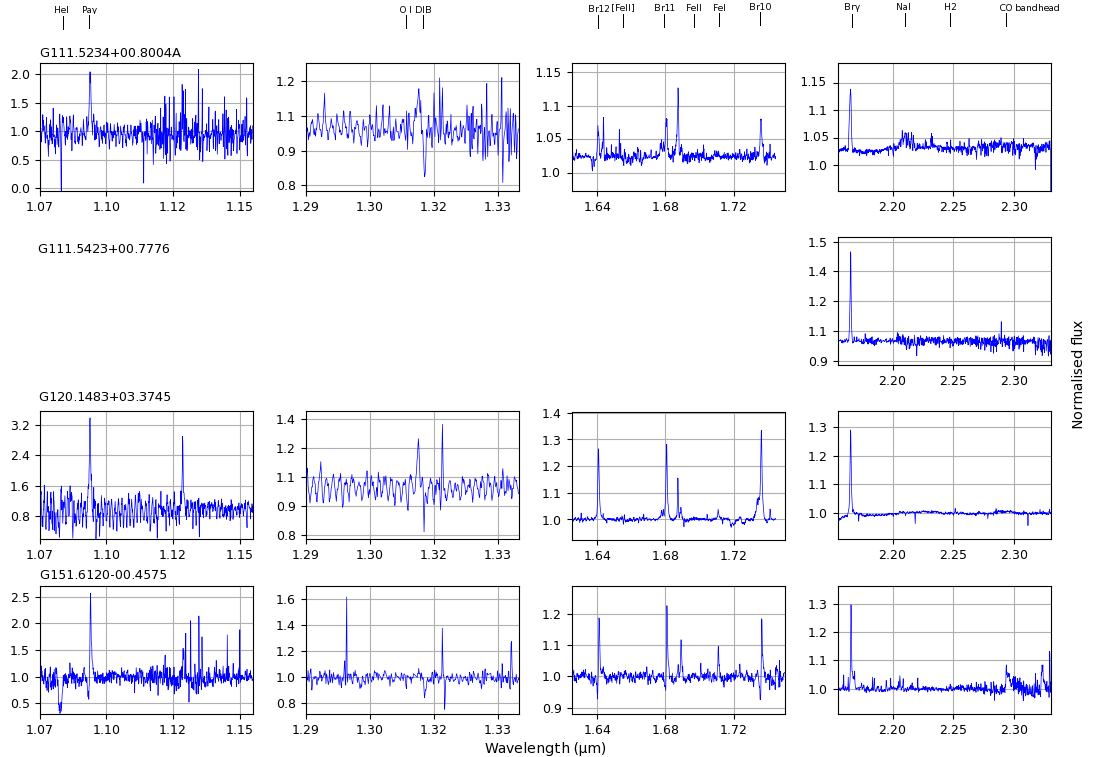}
\caption{Continuum normalised spectra, with left to right panels showing X, J, H and K bands}
\label{fig:sp9}
\end{figure}
\clearpage 
 
\begin{figure}
\includegraphics[scale=0.6]{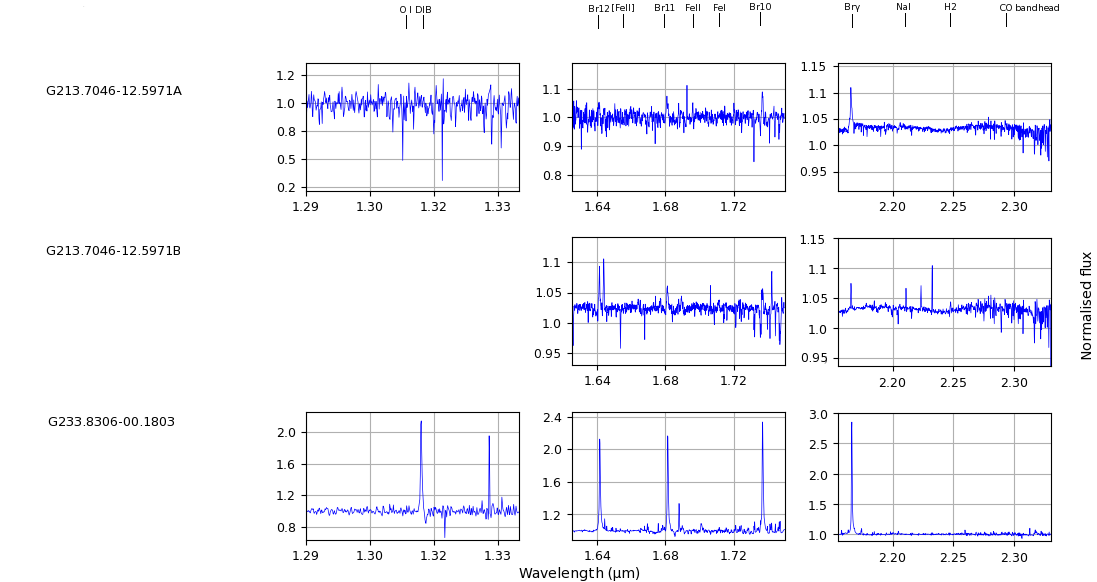}
\caption{Continuum normalised spectra, with left to right panels showing X, J, H and K bands}
\label{fig:sp10}
\end{figure}

\label{lastpage}
\end{landscape}
\end{document}